\documentclass{article}

\PassOptionsToPackage{numbers}{natbib}

\usepackage{amsmath,amsfonts,bm}

\def\eqref#1{equation~\ref{#1}}

\def\1{\bm{1}}

\def\rvx{{\mathbf{x}}}

\def\rvz{{\mathbf{z}}}

\DeclareMathAlphabet{\mathsfit}{\encodingdefault}{\sfdefault}{m}{sl}
\SetMathAlphabet{\mathsfit}{bold}{\encodingdefault}{\sfdefault}{bx}{n}

\def\gO{{\mathcal{O}}}

\def\gT{{\mathcal{T}}}

\def\gV{{\mathcal{V}}}

\usepackage[pagebackref=true,hyperfootnotes=false,breaklinks=true,colorlinks=true,bookmarks=false,linkcolor={red!80!black},urlcolor={darkgray!80!black},citecolor={green!80!black}]{hyperref} 
\usepackage[eandd, final]{neurips_2026}
\usepackage{tikz}
\usepackage{booktabs}
\usepackage{multirow}
\usepackage{adjustbox}
\usepackage{amsmath}
\usepackage{float}

\usepackage[utf8]{inputenc} 
\ifdefined\DeclareUnicodeCharacter
\DeclareUnicodeCharacter{2009}{\,}
\DeclareUnicodeCharacter{202F}{\,}
\fi
\usepackage[T1]{fontenc}    
\usepackage{hyperref}       
\usepackage{url}            
\usepackage{booktabs}       
\usepackage{amsfonts}       
\usepackage{nicefrac}       
\usepackage{microtype}      
\usepackage{xcolor}         
\usepackage{caption}
\usepackage{booktabs}
\usepackage{multirow}

\title{COSMIO: A Benchmark for Cross-Survey Modality Imputation}

\author{%
  \textbf{Dichang Zhang}$^{1}$\thanks{Work done during internships at Futurewei Technologies.}
  \quad \textbf{Yixuan Shao}$^{2}$ \quad \textbf{Jiaqi Deng}$^{3}$ \\
  \textbf{Jiali Cui}$^{4}$ \quad \textbf{Haotian Yin}$^{5,*}$
  \quad \textbf{Yuanpeng Liu}$^{1,*}$ \\
  \textbf{Xiang Gao}$^{1,*}$ \quad \textbf{Zhiqiang Lao}$^{4}$
  \quad \textbf{Heather Yu}$^{4}$\thanks{Corresponding Author} \\
  \textbf{Simon Birrer}$^{2}$ \quad \textbf{Dimitris Samaras}$^{1}$ \\[0.5em]
  $^{1}$Department of Computer Science, Stony Brook University, USA \\
  $^{2}$Department of Physics and Astronomy, Stony Brook University, USA \\
  $^{3}$School of Computer Science, University of Technology Sydney, Australia \\
  $^{4}$Futurewei Technologies, USA \\
  $^{5}$Department of Computer Science, New Jersey Institute of Technology, USA \\[0.5em]
  \texttt{\{diczhang,yuanpliu,gao2,samaras\}@cs.stonybrook.edu} \\
  \texttt{\{yixuan.shao,simon.birrer\}@stonybrook.edu} \\
  \texttt{\{jcui2,zlao,hyu\}@futurewei.com} \\
  \texttt{hy9@njit.edu} \quad \texttt{jiaqi.deng@student.uts.edu.au}
}

\begin{document}

\maketitle

\begin{abstract}
Large-scale astronomical surveys define a heterogeneous, partially observed multi-survey learning setting: each object is observed by only a subset of surveys, while survey observations differ in resolutions, wavelength coverage, noise characteristics, and footprints. This setting gives rise to a natural machine learning problem: given observations of an object in some surveys, how to infer its appearance in other surveys.
We formalize this as the problem of \textbf{cross-survey modality imputation} and introduce \textbf{COSMIO}, a multi-survey benchmark of aligned observations obtained by forward-rendered shared astrophysical scenes across \textit{LSST}, \textit{Euclid}, and \textit{Roman} under survey-specific instrument characteristics.
The benchmark enables systematic evaluation of imputation across source-target survey combinations and missing-survey patterns. We evaluate a broad range of machine learning methods for imputation. 
Our results reveal a clear domain gap: methods developed for existing imputation settings transfer poorly to the astronomical survey setting, indicating that cross-survey imputation constitutes a distinct learning problem.
Beyond common-object imputation, we identify missing-modality rare-event imputation as a central challenge. This regime arises in the early stages of new surveys, when rare but scientifically valuable phenomena may be available only from prior surveys. We use strong gravitational lensing, where light from a background galaxy is gravitationally deflected by a foreground galaxy, as a representative rare-event setting. We evaluate model performance on zero-shot imputation of gravitational lenses and on downstream scientific tasks using the imputed lens observations.
On our constructed evaluation set of galaxy–galaxy strong lenses, performance degrades under rare-event distribution shift, and this degradation further propagates to the downstream lens-detection task.
Our results establish cross-survey imputation as a distinct and challenging machine learning problem, highlighting the need for methods that are robust to distribution shift and aligned with scientific objectives. Code repository: \url{https://github.com/fw-ic/COSMIO}, and the dataset is available at \url{https://www.kaggle.com/datasets/cosmidata/cosmio}.
\end{abstract}
\begin{figure*}[t]
    \centering
    \includegraphics[height=7.2cm,width=\textwidth]{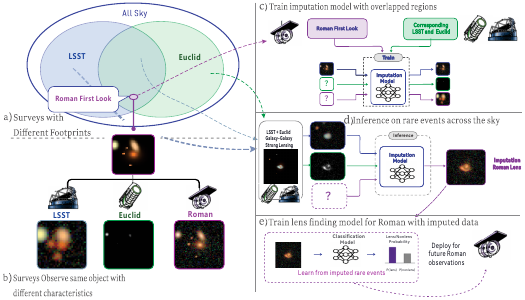}
    \caption{Scientific context for cross-survey rare-event imputation for early-stage observations.
(a) Astronomical surveys cover partially overlapping regions of the sky.
(b) The same underlying astrophysical object is observed with heterogeneous characteristics across surveys.
(c) An imputation model is trained on aligned observations of common galaxies in the overlapping regions.
(d) The trained model is applied to infer missing observations for rare events.
(e) The reconstructed observations enable downstream scientific tasks.}
    
    \label{fig:teaser}
\end{figure*}
\section{Introduction}
Modern astronomical research is increasingly driven by large-scale sky surveys, which observe the Universe with different resolutions, wavelength coverage, noise characteristics, and sky footprints. Flagship surveys such as the Legacy Survey of Space and Time (LSST) at the Vera C. Rubin Observatory~\cite{ivezic2019lsst}, the ESA Euclid mission~\cite{mellier2025euclid}, and upcoming surveys including the Nancy Grace Roman Space Telescope~\cite{roman}, the Simons Observatory~\cite{simon}, and SPHEREx~\cite{spherex} provide complementary measurements under substantially different instrumental conditions. Because their footprints overlap only partially, many objects are available in only a subset of surveys, creating a heterogeneous, partially observed multi-survey setting.

This setting gives rise to a natural machine learning problem. Incomplete cross-survey coverage at the object level limits analyses that require matched observations across surveys. More formally, given observations of an object in some surveys, the goal is to infer how the same object would appear in another survey in which it is unavailable. Such a capability would support more unified analysis across complementary surveys despite incomplete coverage. The problem is particularly consequential during the initial observation phase of a new survey, when only a limited region of the sky has been covered. Under this condition, rare yet scientifically important phenomena, including galaxy--galaxy strong gravitational lenses, mergers and interactions, supernovae and other transients, and unusual morphological outliers such as ring galaxies or tidal features~\cite{Shajib:2024,conselice2014evolution,appleton1996collisional,walmsley2022galaxy}, are typically absent from the target survey, limiting event-specific model development and validation. Predictions from imputation models trained on overlapped regions can serve as proxy target-survey observations for early-stage analysis and method development. Figure~\ref{fig:teaser} illustrates the scientific context.

We formalize this setting as cross-survey modality imputation, where the task is to infer the appearance of the same astrophysical source in a target survey from its available observations in one or more source surveys. 
To study this problem, we introduce \textbf{COSMIO}, a benchmark of forward-rendered, aligned multi-survey observations. COSMIO contains 115k regular galaxy samples and 14{,}825 galaxy--galaxy strong-lensing systems, rendered under realistic LSST, Euclid, and Roman survey configurations. 



We use COSMIO to study three questions: (i) can models infer missing survey observations from available surveys? (ii) can models trained on common objects generalize to scientifically important rare-event regimes? and (iii) can imputed observations improve downstream scientific tasks?
Our empirical study yields three main findings. First, methods developed for the most established missing-modality imputation setting, namely multi-contrast brain MRI imputation~\cite{wu2026codebrain,diffm4ri,mmgan,ummi,mmsyn,m2dn}, transfer poorly to astronomical survey data and are often outperformed by simple diffusion baselines, revealing a clear domain gap. Second, models trained on common galaxies degrade when evaluated on rare galaxy--galaxy strong-lensing systems. Third, imputed observations can support downstream strong-lens detection, but their performance remains below that of real target-survey observations and depends strongly on reconstruction quality.
Together, these findings establish cross-survey imputation as a distinct modality-imputation setting and highlight the need for dedicated methods for multi-survey astronomical data, a problem that remains unsolved.




Our contributions are as follows: 
(\textbf{i}) We formulate cross-survey modality imputation as a new learning problem setting arising from partially overlapping astronomical surveys, and define the associated scientific setting and machine learning tasks. (\textbf{ii}) We introduce COSMIO, a benchmark with 115k regular galaxy samples and 14{,}825 strong-lensing systems across LSST, Euclid, and Roman, enabling systematic evaluation under realistic cross-survey configurations. (\textbf{iii}) We show that methods developed for existing imputation tasks transfer poorly to this regime and identify rare-event imputation as a key challenge which affects downstream scientific applications. 



\section{Related Work}

\subsection{Next-Generation Sky Surveys}

Modern observational cosmology is entering a multi-survey era driven by wide-field facilities such as the Legacy Survey of Space and Time (LSST) at the Vera C. Rubin Observatory~\cite{ivezic2019lsst}, the ESA Euclid mission~\cite{mellier2025euclid}, SPHEREx~\cite{spherex} and upcoming facilities including the Nancy Grace Roman Space Telescope~\cite{roman}, the Simons Observatory~\cite{simon}. These surveys observe overlapping astrophysical populations, but through substantially different instruments and observing conditions. In this work, we focus on three representative surveys, including LSST, Euclid, and Roman, which form a prototypical multi-survey setting with substantial sky coverage and strong complementarity across resolution, depth, and spectral regimes. 

\noindent\textbf{Legacy Survey of Space and Time~\cite{ivezic2019lsst}}. LSST is a ground-based, wide-field optical survey designed for high-cadence, deep imaging over a large fraction of the sky. It operates in wide optical bands and repeatedly scans the sky over a ten-year period. The survey is expected to begin full science operations soon this year (following first light in 2025), with incremental data releases throughout the survey and a final data release at the end of its ten-year program. LSST will cover approximately 18,000 deg² of the southern sky, forming one of the largest and deepest optical survey footprints. However, as a ground-based instrument, LSST observations are affected by atmospheric seeing, leading to PSF-dominated resolution limitations and variable noise characteristics.

\noindent\textbf{ESA Euclid Mission~\cite{mellier2025euclid}.} Euclid is a space-based mission focused on high-precision cosmology through wide-area imaging and spectroscopy. Launched in 2023, Euclid is currently in its operational phase, with early public data products released, while the full wide survey is still ongoing. It aims to survey roughly 15,000 deg² of extragalactic sky, avoiding regions of high stellar density near the Galactic plane. Compared to LSST, Euclid offers significantly sharper morphology. However, its observational design trades off spectral richness and cadence, resulting in limited band coverage.

\noindent\textbf{Nancy Grace Roman Space Telescope.~\cite{roman}} Roman extends this landscape with wide-field, space-based near-infrared imaging and spectroscopy. Roman is scheduled for launch in Fall 2026, with major survey programs expected to begin shortly thereafter. Its planned surveys will cover smaller areas (typically thousands of deg² depending on survey strategy) but at significantly greater depth. 

Table~\ref{tab:survey_configs} summarizes the key observational properties of these surveys.

\begin{table}[t]
\caption{Comparison of key observational parameters across LSST, Euclid, and Roman surveys.}
\centering
\small
\label{tab:survey_configs}
\begin{tabular}{lccc}
\toprule
Parameter & LSST & Euclid & Roman \\
\midrule
Bands 
& $g, r, i, z, y$ 
& VIS 
& $F106(Y), F129(J), F158(H)$ \\

PSF (arcsec) 
& $\sim 0.7''$ 
& $\sim 0.16''$ 
& $\sim 0.11''$ \\

Pixel Scale (arcsec) 
& $0.2''$ 
& $0.1''$ 
& $0.11''$ \\

Depth 
& $r \sim 27.5$ 
& VIS $\sim 24.5$ 
& $H \sim 26$--$27$ \\

Noise 
& seeing $+$ sky 
& readout 
& thermal $+$ readout \\

Survey Area (deg$^2$) 
& $\sim 18{,}000$ 
& $\sim 15{,}000$ 
& $\sim 2{,}000$--$5{,}000$ \\
\bottomrule
\end{tabular}

\label{tab:survey_comparison}
\end{table}

\subsection{Image Restoration in Astronomy}
Recent learning-based methods have explored the use of deep generative models for astronomical settings. For example, GAN-based methods have been used to recover galaxy structures ~\cite{schawinski2017generative}, and diffusion models have been applied to super-resolution of gravitational lensing images~\cite{reddy2024difflense}. Several benchmarks have also been proposed for astronomical super-resolution, including AstroSR for galaxy images~\cite{miao2024astrosr} and STAR for flux-consistent star-field super-resolution~\cite{wu2026star}. However, these works focus on observations from a single telescope with a single modality. 

Another line of work reconstructs images or videos from sparse interferometric measurements collected by telescope arrays~\cite{bouman2016computational,bouman2017reconstructing}. These works involve multiple instruments, but the instruments belong to the same observational system and share the same measurement modality, differing mainly in spatial or temporal sampling.
\subsection{Missing Modality Imputation}

Missing-modality imputation studies how to infer unobserved measurements of the same underlying object from its available measurements under other acquisition processes. 
With an unobserved latent physical state $\rvz$, the $k$-th modality is an instrument-specific realization
\begin{equation}
    \rvx^{(k)} = \mathcal{A}_k(\rvz) + \epsilon_k,
\end{equation}
where $\mathcal{A}_k$ denotes the acquisition function of the $k$-th survey or instrument, and $\epsilon_k$ denotes the corresponding observational noise.

Formally, let $\gV=\{1,\dots,K\}$ denote the set of available modalities, and let
$\rvx=\left(\rvx^{(k)}\right)_{k\in\gV}$ be the collection of observations of the same underlying object under these acquisition functions. 
Given an observed subset $\gO \subset \gV$ and a target subset $\gT\subseteq (\gV\setminus\gO)$, the imputation task is to model the conditional distribution
\begin{equation}
    p_\theta\left(\rvx^{(\gT)} \mid \rvx^{(\gO)}\right),
\end{equation}
where $\rvx^{(\gO)}=\{\rvx^{(k)}:k\in\gO\}$ and $\rvx^{(\gT)}=\{\rvx^{(k)}:k\in\gT\}$.

This problem has been extensively studied for multi-contrast brain MRI, where different image types, including T1, T2, FLAIR, and contrast-enhanced T1 MRI, observe the same brain under different acquisition contrasts. Each modality emphasizes different tissue properties, anatomical structures, or pathological patterns, and their combination provides a more complete characterization of the same subject than any single MRI sequence alone. Recent brain MRI imputation methods include CodeBrain~\cite{wu2026codebrain}, DiffM4RI~\cite{diffm4ri}, M2DN~\cite{m2dn}, MMGAN~\cite{mmgan}, MMSyn~\cite{mmsyn}, and UMMI~\cite{ummi}, and are commonly evaluated on BraTS benchmarks~\cite{menze2015brats,bakas2018identifying}.

Our astronomical surveys share a high-level similarity with medical imaging in that multiple observations can be viewed as different measurements of the same underlying object. However, the main difference lies in the relation between observed and target modalities. MRI contrasts are typically acquired on a common scanner with a common imaging geometry under relatively controlled conditions, whereas astronomical surveys differ across instruments and observation conditions, leading to larger variations in spatial resolution, point spread functions, wavelength coverage and noise characteristics. Our experiments show that the unified multi-to-multi formulation remains useful in the survey setting, whereas MRI-specific designs transfer poorly. This difference motivates cross-survey modality imputation as a separate benchmark problem rather than a direct extension of existing MRI imputation methods.

\section{COSMIO}
\label{sec:cosmio}

COSMIO is constructed as a controlled multi-survey observation benchmark. We
first define an underlying astrophysical scene and then render how that same
scene would be observed by LSST, Euclid, and Roman. 

Formally, let $s$ denote a latent sky scene. A scene contains the intrinsic
astronomical content before observation, such as galaxy type, redshift,
brightness, morphology, spatial position, and, for lensing systems, the
foreground--background geometry that determines light deflection. In COSMIO,
these source-level properties are sampled from real galaxy catalogs and then
forward-rendered under survey-specific observing conditions. Each survey
image is rendered by applying a survey-specific observation operator, $x^{(k)} = \mathcal{O}_k(s) + \epsilon_k$,
where $k \in \{\mathrm{LSST}, \mathrm{Euclid}, \mathrm{Roman}\}$,
$\mathcal{O}_k$ represents the deterministic part of the imaging process, and
$\epsilon_k$ represents survey-specific stochastic noise. The operator
$\mathcal{O}_k$ includes the instrument bandpasses, angular resolution, point spread function (PSF),
pixelization, and exposure configuration.

We generate two types of scenes. The first type consists of common galaxy
fields. These are ordinary galaxy observations sampled from catalog-level
population models. Each galaxy is assigned physically meaningful attributes,
including redshift, spectral energy distribution, luminosity, size, ellipticity,
and surface-brightness profile. To avoid an overly simplified single-object
setting, common-galaxy scenes may also include projected neighboring galaxies.
These additional field galaxies are sampled from the same population model and
placed at random sky positions. They introduce realistic crowding and blending astronomical scenes.

The second type consists of rare galaxy--galaxy strong-lensing systems. These
scenes are generated by pairing a foreground deflector galaxy with a more
distant background source galaxy. The deflector is assigned a mass distribution,
and the background source is assigned a light profile in the source plane. The
lensing model then solves the lens equation to map source-plane coordinates to
image-plane coordinates. This mapping determines where the source light appears
after deflection by the foreground mass. Depending on the relative alignment,
deflector mass, and source morphology, the same background source can be
rendered as multiple distorted images, extended arcs, or Einstein-ring-like
structures~\cite{Shajib:2024}. Thus, the arc morphology is not imposed as
an image template; it is produced by the physical lensing geometry and the
sampled source and deflector properties. An introduction to galaxy--galaxy
strong lensing is provided in Appendix~\ref{app:strong-lensing}. We retain only
lens configurations that are sufficiently resolved and bright to serve as
meaningful rare-event examples. As in the common-galaxy scenes, lensing scenes
may also include projected field galaxies, which are rendered as additional
light components at random sky positions.

To preserve survey-native angular sampling while maintaining an approximately
consistent field of view across surveys, we choose survey-specific cutout sizes
based on the native pixel scale of each instrument instead of enforcing a common
pixel grid. In the current dataset release, we use cutout sizes of
$64\times64$ for Euclid, $32\times32$ for LSST, and $59\times59$ for Roman,
yielding approximately matched angular coverage across the three surveys while
retaining their native sampling characteristics.

\begin{figure*}[!t]
    \centering
    \includegraphics[width=\textwidth]{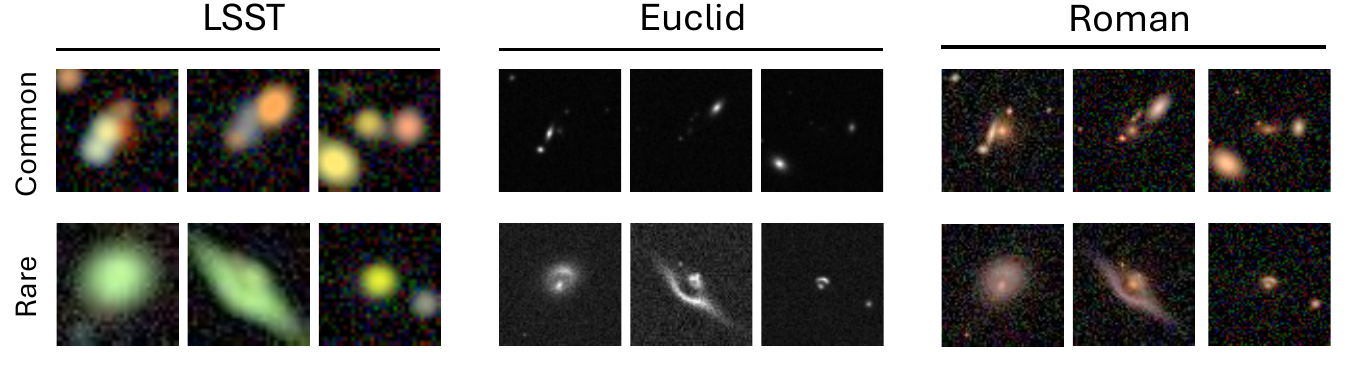}
    \caption{COSMIO samples across three survey modalities, including both common galaxy fields and rare scenes such as strong-lensing systems. Additional dataset visualizations are provided in Appendix~\ref{app:dataset_visualizations}.}
    \label{fig:examples}
\end{figure*}

In total, COSMIO contains 115{,}000 common-galaxy scenes and 14{,}825 scenes
containing strong lensing systems. Figure~\ref{fig:examples} shows some sample data from the dataset.  Implementation details, including the software packages,
cosmological parameters, survey bandpasses, selection cuts, and
data format, are provided in Appendix~\ref{app:data-gen}.

\section{Experiments}

\subsection{Baseline Methods}

We compare a range of methods from the medical image domain, diffusion baselines, and two physically motivated baselines. For medical-domain baselines, we use CodeBrain ~\cite{wu2026codebrain} and include DiffM4RI~\cite{diffm4ri}, MM-GAN~\cite{mmgan}, UMMI~\cite{ummi}, MMSyn~\cite{mmsyn}, and M2DN~\cite{m2dn}. Full descriptions are provided in Appendix~\ref{app:mri-architecture-details}.

Except for imputation methods adapted from medical image problem, we further evaluate three diffusion baseline. The first is Palette~\cite{saharia2022palette}, where we train a separate diffusion model for each imputation pattern. The model is conditioned on the observed survey modalities and learns to generate the missing modality, approximating the conditional distribution of the target observation given the available inputs. 

Beyond this straightforward baseline, we consider two unified diffusion variants. This design is motivated by prior evidence from medical image imputation that a single model trained across multiple missing-modality patterns often generalizes better than pattern-specific models.

The second diffusion baseline is RePaint~\cite{repaint}. RePaint models the full joint distribution over all modalities using a single model. During sampling, at each reverse diffusion step, the observed modalities are first noised to the current timestep using the forward process, and then used to replace the corresponding part of the sample, while the missing modalities are updated by the model. This ensures consistency with the observations while allowing the model to generate the missing modalities.

We further propose an asynchronous diffusion baseline.
Our goal with Asyn-Diff is primarily subtraction rather than architectural addition: removing MRI-specific components to establish a lightweight, reproducible reference baseline for cross-survey imputation. Relative to RePaint's original spatial-inpainting formulation, Asyn-Diff uses a modality-level mask, treating each survey modality as entirely observed or entirely missing. Relative to M2DN, it removes separate modality embeddings and specialized modules such as channel attention. Modality availability is encoded directly through modality-specific timestep inputs to a minimal three-level U-Net. Relative to our Palette baseline, which trains separate conditional models for different imputation patterns, Asyn-Diff jointly trains all modalities and imputation patterns in a single model. The contribution is this deliberately minimal modality-level formulation as a strong reference baseline, rather than a fundamentally new diffusion mechanism. Architectural details are provided in Appendix~\ref{app:diffusion-baselines}.

We also include two Physically motivated baselines.
We include PSF-Linear, which combines PSF homogenization with a global ridge-regression color transform, and S\'ersic-FM, which jointly fits a shared S\'ersic morphology to the input bands and forward-renders native-resolution Roman predictions. Full descriptions are provided in Appendix~\ref{app:physical-baselines}.

\subsection{Experimental Setup}
\label{sec:experimental-setup}

\paragraph{Training details.}

For the neural baselines, we normalize each survey band using a signed, capped $asinh$ transform fitted independently for that band. The transform estimates a robust center and scale and clips only the most extreme percentile tails, preserving the wide dynamic range of astronomical images while mapping Euclid, LSST, and Roman pixel values to a common bounded range for stable model training. We also pad all corresponding LSST, Euclid, and Roman observations to aligned $64\times64$ patches, rather than resampling them to a common pixel scale. This choice preserves each survey’s native pixel grid and telescope-specific receptive field; rescaling would mix neighboring pixels and distort the approximately independent pixel-noise structure induced by the detector and observation process.

For neural imputation models other than CodeBrain, we train on regular galaxies. Optimization is performed with AdamW at a learning rate of $1\times10^{-4}$. We use a batch size of 128 and train for 100 epochs on a single NVIDIA H100 GPU. All diffusion models use $1,000$ timesteps, a linear beta schedule from $1\times10^{-4}$ to $2\times10^{-2}$, epsilon prediction, and clipped $x_0$ estimates.
For downstream classification, we train ViT models under different data source settings. The training data are constructed from: (1) Roman images from forward physical modeling (GT); (2) Roman images imputed from LSST and Euclid jointly; (3) Roman images imputed from Euclid only; and (4) Roman images imputed from LSST only.
All ViT classifiers are trained for 10 epochs with a batch size of 64. Training is performed with cross-entropy loss and balanced sampling over the training split to mitigate class imbalance.

For the main zero-shot comparison, the neural imputation models are trained exclusively on common galaxies, with 90,000 training examples and 9,000 validation examples. Imputation quality is evaluated on a held-out downstream-training split containing 7,481 strong-lensing systems and 8,000 common galaxies. We use the corresponding imputed observations from this split to train downstream lens classifiers, and evaluate classification performance on another disjoint downstream-test split containing 7,344 strong-lensing systems and 8,000 common galaxies.

\providecommand{\second}[1]{\underline{#1}}

\begin{table*}[t]
\centering
\small
\caption{Performance comparison of cross-survey imputation methods on common galaxies.
Results are grouped by target survey, with metrics averaged over all corresponding imputation tasks. Best results are in \textbf{bold}, and second-best results are \second{underlined}. CRPS is not reported for the two deterministic physical baselines. Detailed results for individual Roman-target imputation tasks are provided in Appendix~\ref{app:detailed-tables}.}
\label{tab:target_avg_results}
\begin{tabular}{llccc}
\toprule
Tasks & Method & RMSE$\downarrow$ & PSNR$\uparrow$ & CRPS$\downarrow$ \\
\midrule

\multirow{11}{*}{%
\begin{tabular}[c]{@{}l@{}}

$\text{Euclid} \rightarrow \text{Roman}$\\
$\text{LSST} \rightarrow \text{Roman}$\\
$\text{Euclid+LSST} \rightarrow \text{Roman}$\\
\end{tabular}}
& PSF-Linear & 0.403 & 33.758 & / \\
& S\'ersic-FM & 0.271 & 36.838 & / \\
& CodeBrain~\cite{wu2026codebrain} & 0.330 & 34.533 & 0.097 \\
& diffm4ri~\cite{diffm4ri}   & 0.353 & 34.153 & 0.094 \\
& mmgan~\cite{mmgan}      & 0.522 & 30.157 & 0.159 \\
& ummi~\cite{ummi}       & 0.449 & 33.567 & 0.134 \\
& mmsyn~\cite{mmsyn}      & 0.282 & 35.727 & 0.092 \\
& m2dn~\cite{m2dn}       & 0.238 & 37.423 & 0.077 \\
& RePaint~\cite{repaint}    & 0.322 & 35.507 & 0.090 \\
& Palette~\cite{saharia2022palette}    & \second{0.196} & \second{38.737} & \second{0.072} \\
& asyn-diff  & \textbf{0.184} & \textbf{39.190} & \textbf{0.069} \\

\midrule

\multirow{11}{*}{Euclid $\rightarrow$ LSST}
& PSF-Linear & \second{0.285} & \second{33.170} & / \\
& S\'ersic-FM & 0.336 & 31.571 & / \\
& CodeBrain~\cite{wu2026codebrain} & \textbf{0.241} & \textbf{34.580} & \second{0.112} \\
& diffm4ri~\cite{diffm4ri}   & 3.327 & 9.420  & 1.407 \\
& mmgan~\cite{mmgan}      & 1.418 & 20.330 & 0.378 \\
& ummi~\cite{ummi}       & 1.161 & 24.010 & 0.374 \\
& mmsyn~\cite{mmsyn}      & 0.393 & 32.330 & 0.161 \\
& m2dn~\cite{m2dn}       & 0.314 & 32.420 & 0.137 \\
& RePaint~\cite{repaint}    & 0.953 & 25.720 & 0.291 \\
& Palette~\cite{saharia2022palette}    & 0.870 & 24.250 & 0.231 \\
& asyn-diff  & 0.332 & 32.690 & \textbf{0.111} \\

\midrule

\multirow{11}{*}{LSST $\rightarrow$ Euclid}
& PSF-Linear & 0.017 & 29.217 & / \\
& S\'ersic-FM & 0.011 & 32.948 & / \\
& CodeBrain~\cite{wu2026codebrain}& 0.011 & 32.200 & 0.006 \\
& diffm4ri~\cite{diffm4ri}   & 0.014 & 30.430 & 0.005 \\
& mmgan~\cite{mmgan}      & 0.030 & 23.070 & 0.010 \\
& ummi~\cite{ummi}       & 0.021 & 28.940 & 0.008 \\
& mmsyn~\cite{mmsyn}      & 0.011 & 32.180 & 0.006 \\
& m2dn~\cite{m2dn}       & 0.010 & 33.330 & 0.005 \\
& RePaint~\cite{repaint}    & 0.014 & 31.020 & 0.005 \\
& Palette~\cite{saharia2022palette}    & \second{0.009} & \second{33.740} & \second{0.004} \\
& asyn-diff  & \textbf{0.009} & \textbf{33.820} & \textbf{0.004} \\

\bottomrule
\end{tabular}
\end{table*}

\paragraph{Metric hierarchy and evaluation protocols.}
We use the Continuous Ranked Probability Score (CRPS)~\cite{Gneiting01032007} as the primary general reconstruction metric for probabilistic methods. For each probabilistic method and test input, we draw five stochastic samples and evaluate
$\mathrm{CRPS}(\hat{F},y)=\mathbb{E}_{\hat{Y}\sim\hat{F}}|\hat{Y}-y|-\frac{1}{2}\mathbb{E}_{\hat{Y},\hat{Y}'\sim\hat{F}}|\hat{Y}-\hat{Y}'|$.
CRPS balances proximity to the observation with predictive dispersion. Its expected value is uniquely minimized by the true distribution of the evaluated scalar quantity, rather than by a collapsed point prediction when that distribution is non-degenerate. This motivates its use for uncertain cross-survey predictions. 

RMSE and PSNR are secondary, per-pixel reconstruction diagnostics, computed from raw-flux images; PSNR uses the maximum ground-truth pixel value of each observation as its peak value. 

We further report physically motivated diagnostics in Appendix~\ref{app:scientific-diagnostics}. Although scientifically informative, these measurements use task-specific definitions and extraction procedures rather than a standardized evaluation protocol. We therefore treat them as reference diagnostics. Recovery of low-dimensional summary statistics is also less central to COSMIO's intended use in reconstructing spatially resolved morphology. 

Downstream lens classification, measured by AUROC and AUPRC, directly evaluates the utility of the imputed images for our demonstrated scientific task.

We do not adopt SSIM or perceptual metrics such as LPIPS~\cite{lpips} as primary evaluation criteria. SSIM measures image-space structural similarity, while LPIPS relies on natural-image feature representations; neither directly captures the scientific fidelity required for downstream tasks.

\subsection{Results}
\label{sec:results}
\paragraph{Regular Galaxies} We first compare the performance of baseline methods on imputing regular galaxies. The results are reported in Table~\ref{tab:target_avg_results}. Tasks are defined as all possible combinations of input and output modalities, and are grouped by target survey. 
Asynchronous diffusion has the lowest reported CRPS in each target-survey group. Both Asynchronous diffusion and conditional diffusion consistently outperform medical imputation baselines. This suggests that cross-survey modality imputation differs fundamentally from MRI imputation, and that architecture choices tailored to medical imaging do not generalize well to the astronomical setting.

\begin{center}
    \begin{minipage}{\linewidth}
        \label{tab:lens_flux_results}
        \captionof{table}{Average performance comparison of cross-survey imputation methods on rare galaxy--galaxy strong lensing.
        Results are averaged over three Roman-imputation tasks. Best results are in bold, and second-best results are underlined. CRPS is not reported for the two deterministic physical baselines. Detailed results for individual Roman-target imputation tasks are provided in Appendix~\ref{app:detailed-tables}. AEE is reported separately in Appendix~\ref{app:angular-extent-error}. }
        \centering
        \small
        \label{tab:lens_flux_results_avg}
        \begin{tabular}{llccc}
        \toprule
        Tasks & Method & RMSE$\downarrow$ & PSNR$\uparrow$ & CRPS$\downarrow$ \\
        \midrule
        \multirow{7}{*}{%
        \begin{tabular}[c]{@{}l@{}}
        
        $\text{Euclid} \rightarrow \text{Roman}$\\
        $\text{LSST} \rightarrow \text{Roman}$\\
        $\text{Euclid+LSST} \rightarrow \text{Roman}$\\
        
        \end{tabular}}
& PSF-Linear & 0.572 & 32.894 & / \\
& S\'ersic-FM & 0.538 & 33.759 & / \\
& CodeBrain\cite{wu2026codebrain} & 0.440 & 34.127 & 0.124 \\
        & m2dn~\cite{m2dn}      & 0.356 & 36.360 & 0.106 \\
        & RePaint~\cite{repaint}   & 0.522 & 34.040 & 0.144 \\
        & Palette~\cite{saharia2022palette}   & \second{0.310} & \second{37.163} & \second{0.095} \\
        & asyn-diff & \textbf{0.297} & \textbf{37.527} & \textbf{0.090} \\
        \bottomrule
        \end{tabular}
    \end{minipage}
\end{center}

\begin{center}
    \begin{minipage}{\linewidth}
        \centering
        \includegraphics[width=\textwidth]{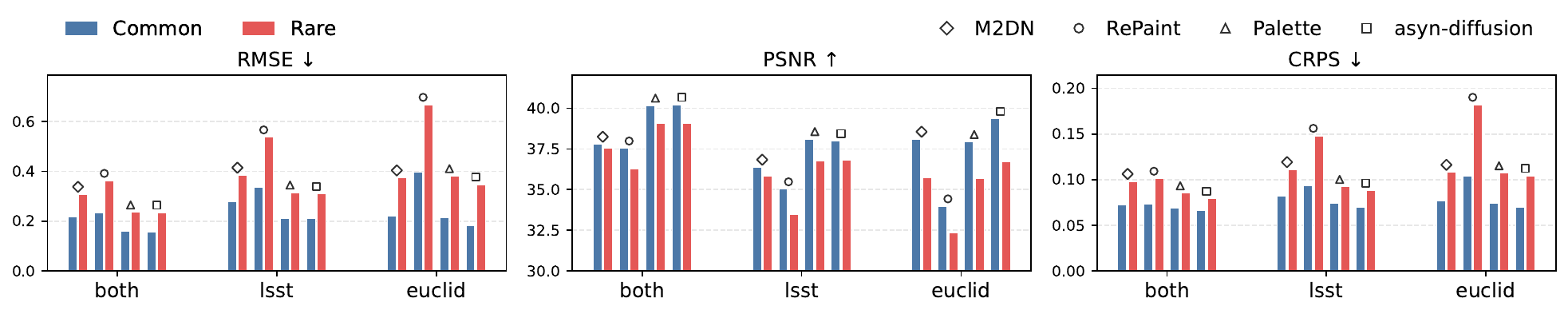}
        \captionof{figure}{Performance of Roman imputation from LSST, Euclid, or both under distribution shift from common galaxies to rare events. Performance degrades consistently across all three metrics, three imputation patterns, and four methods.}
        
        \label{fig:regular2rare}
    \end{minipage}
\end{center}

\paragraph{Rare Events}
We futher rare-event imputation results measure \textbf{zero-shot transfer from common galaxies to strong-lensing systems}: the imputation models receive no lens-target supervision before this evaluation. We compare the two physical baselines, all diffusion baselines, and the MRI imputation baselines M2DN and CodeBrain in Table~\ref{tab:lens_flux_results_avg}. Here zero-shot refers to the imputation stage; downstream lens classifiers are trained separately on imputed lens and non-lens images. Appendix~\ref{app:fewshot-mitigation} evaluates lens-only fine-tuning and importance-weighted joint training with limited lens supervision as potential mitigations of this rare-event shift.
Our asynchronous diffusion baseline achieves the lowest average CRPS in Table~\ref{tab:lens_flux_results_avg}.  Its strong performance on both common galaxies and strong-lensing systems makes it a useful reference baseline for future method development in cross-survey imputation. However, as shown in Figure~\ref{fig:regular2rare}, all methods degrade noticeably compared with their performance on regular galaxies, revealing a substantial rare-event distribution shift. Figure~\ref{fig:imputation-examples} illustrates this failure mode. For an in-distribution edge-on disk galaxy, the model preserves the elongated morphology when imputing Roman observations from LSST, Euclid, or both. Although edge-on galaxies are visually distinctive, their flux distribution is still broadly consistent with the centrally concentrated profiles of regular galaxies. By contrast, strong-lensing systems contain extended arc- or ring-like structures caused by lensing, which deviate from the dominant morphology learned from common galaxies. Probabilistic generative imputation models therefore tend to favor high-density target-modality structures when the rare-event evidence is ambiguous, causing the ring to fragment into compact galaxy-like point sources.

\begin{center}
    \begin{minipage}{\textwidth}
    \centering
    \includegraphics[width=\textwidth]{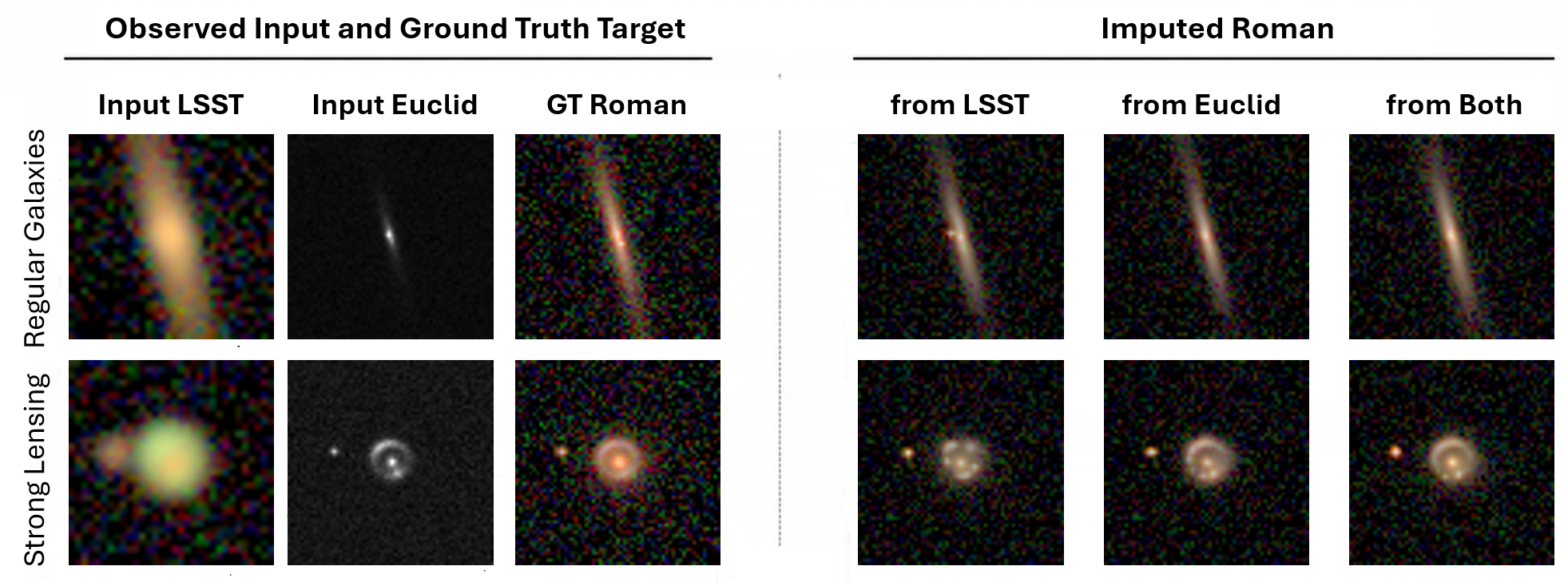}
    \captionof{figure}{Qualitative examples of Roman imputation under different observed modality sets. For an in-distribution edge-on disk galaxy, the model preserves the elongated morphology when imputing Roman from LSST, Euclid, or both. In contrast, for an out-of-distribution strong-lensing system, the ring-like lensing structure is not faithfully preserved and is partially broken into compact sources. More visualizations are provided in Appendix~\ref{app:imputation_visualizations}.}
    \label{fig:imputation-examples}
\end{minipage}
\end{center}

\begin{center}
    \begin{minipage}{\linewidth}
        \centering
        \includegraphics[width=\textwidth]{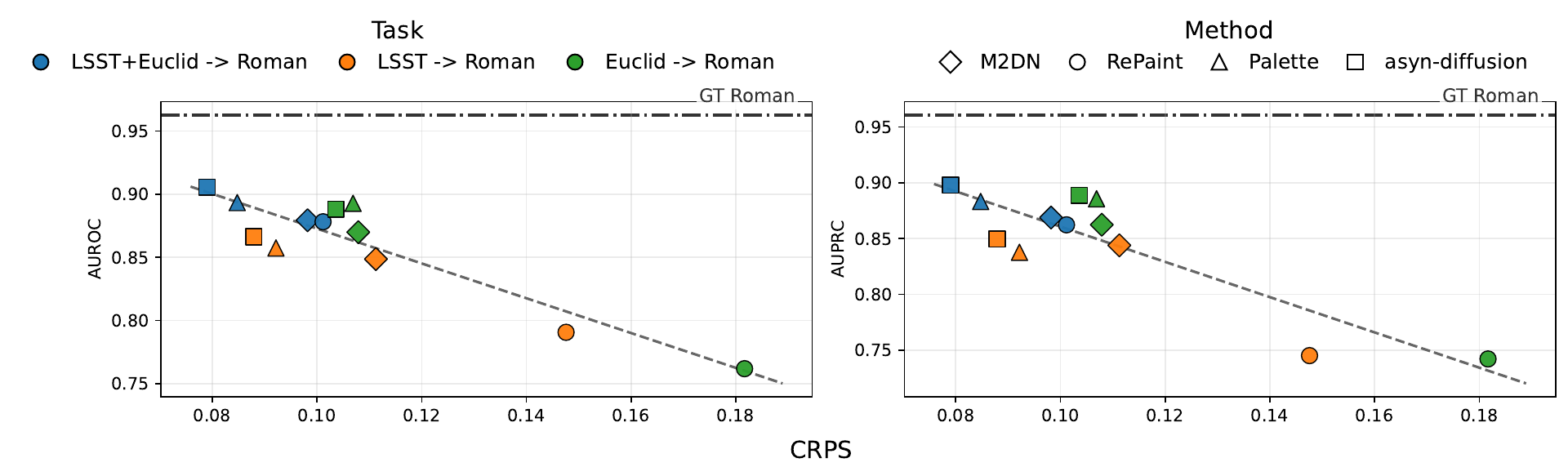}
        \captionof{figure}{Relationship between imputation quality and downstream classification performance.
    Each point represents one method on one survey-imputation task. Horizontal dash-dotted lines denote downstream performance using ground-truth Roman observations, while dashed lines show fitted trends across imputed inputs. Numerical AUROC and AUPRC values are reported in Appendix~\ref{app:detailed-tables}.}
        
        \label{fig:imp-vs-cls}
    \end{minipage}
\end{center}

\paragraph{Downstream Classification} We further analyze how imputation quality translates to downstream scientific performance. Figure~\ref{fig:imp-vs-cls} shows the relationship between imputation quality, measured by CRPS, and downstream lens classification performance, measured by AUROC and AUPRC.
This comparison tests whether gains in probabilistic reconstruction quality translate into improved downstream detection. The classification metrics complement CRPS by measuring scientific utility directly.

\section{Conclusion}
In this work, we introduced COSMIO, a benchmark for cross-survey astronomical modality imputation. Our experiments show that this setting is distinct from conventional imputation domains, as off-the-shelf methods do not readily generalize to astronomical survey data. We identify rare-event imputation as a key bottleneck, with current models degrading on scientifically important but underrepresented phenomena. We further show that imputation quality directly affects downstream scientific tasks, highlighting the need for more robust imputation methods for next-generation astronomical surveys. Finally, we establish asynchronous diffusion as a strong baseline for future method development.

From an astronomical perspective, COSMIO provides a dataset for synthesizing complementary observations across heterogeneous surveys, enabling analysis in partially observed sky regions. From a machine learning perspective, COSMIO extends modality imputation beyond existing domains by using cross-survey astronomy as a testbed where standard missing-modality inference must also remain robust under rare-event distribution shifts. Potential applications beyond the demonstrated lens-detection task include generating candidate realizations before follow-up or Target-of-Opportunity (ToO) observations to support planning and prioritization, and stress-testing early-survey image-analysis pipelines. These uses remain directions for future evaluation.

\section{Limitations and Future Work}
\label{sec:limitation}
Our dataset is constructed through forward physical rendering, which inevitably introduces a sim-to-real gap. Large-scale overlapping observations across next-generation surveys are not yet available, making simulation-based data the only viable approach for controlled method development. This simulation-first paradigm is standard in astronomy, where physically grounded forward models are widely used to approximate observational processes.

Moreover, the rarity of scientifically important phenomena further necessitates simulation. Real surveys contain only a limited number of such events, which is insufficient for training and systematic evaluation.
Even as simulation fidelity improves, discrepancies between simulated and real survey data will inevitably remain. This limits the reliability of method development and calibration when they are based solely on physical simulations. This limitation is one motivation for introducing cross-survey imputation as a complement to pure physical simulation.

Once authorized Rubin access and sufficient overlapping imaging are available, we plan to cross-match Rubin/LSST and Euclid observations, construct disjoint real-data training and test splits, and train and evaluate each method on these real matched pairs. We will compare method rankings, relative performance gaps, and failure modes with those from the corresponding COSMIO LSST--Euclid tasks. This evaluation will test whether conclusions from the controlled benchmark hold on real observations, with training and testing performed within each domain. It will complement COSMIO's three-survey and rare-event evaluations and can be extended to Roman when suitable matched observations become available.

A complementary direction for mitigating the sim-to-real gap is to incorporate observed galaxy morphologies into COSMIO's source population. We plan to adopt the catalog-based source rendering available in newer versions of SLSim~\cite{khadka2026slsim}, using high-resolution HST/COSMOS galaxy cutouts as source templates. These templates can be matched by physical size and axis ratio, optionally including S\'ersic index, then rescaled and rotated to the desired source geometry before forward rendering through each survey's instrument model. This would introduce observed substructure and irregular morphology beyond analytic profiles. We will evaluate whether this extension improves agreement with real-data method rankings and failure modes.

\bibliographystyle{cosmio_unsrt}
\bibliography{references}

@article{Sersic:1963,
  author = {{S{\'e}rsic}, J.~L.},
  title = {{Influence of the atmospheric and instrumental dispersion on the brightness distribution in a galaxy}},
  journal = {Boletín de la Asociación Argentina de Astronomía},
  year = {1963},
  volume = {6},
  pages = {41--43},
  adsurl = {https://ui.adsabs.harvard.edu/abs/1963BAAA....6...41S},
  url = {https://scixplorer.org/abs/1963BAAA....6...41S/abstract},
}

@article{Saha:2024,
  author = {Saha, Prasenjit and Sluse, Dominique and Wagner, Jenny and Williams, Liliya L. R.},
  title = {{Essentials of Strong Gravitational Lensing}},
  journal = {Space Science Reviews},
  year = {2024},
  volume = {220},
  number = {1},
  eid = {12},
  pages = {12},
  doi = {10.1007/s11214-024-01041-w},
  archiveprefix = {arXiv},
  eprint = {2401.04165},
  primaryclass = {astro-ph.CO},
  adsurl = {https://ui.adsabs.harvard.edu/abs/2024SSRv..220...12S},
  url = {https://doi.org/10.1007/s11214-024-01041-w},
}

@article{Vegetti:2024,
  author = {Vegetti, S. and Birrer, S. and Despali, G. and Fassnacht, C. D. and Gilman, D. and Hezaveh, Y. and Perreault Levasseur, L. and McKean, J. P. and Powell, D. M. and O’Riordan, C. M. and Vernardos, G.},
  title = {{Strong Gravitational Lensing as a Probe of Dark Matter}},
  journal = {Space Science Reviews},
  year = {2024},
  volume = {220},
  number = {5},
  eid = {58},
  pages = {58},
  doi = {10.1007/s11214-024-01087-w},
  archiveprefix = {arXiv},
  eprint = {2306.11781},
  primaryclass = {astro-ph.CO},
  adsurl = {https://ui.adsabs.harvard.edu/abs/2024SSRv..220...58V},
  url = {https://doi.org/10.1007/s11214-024-01087-w},
}

@article{Shajib:2024,
  author = {Shajib, A. J. and Vernardos, G. and Collett, T. E. and Motta, V. and Sluse, D. and Williams, L. L. R. and Saha, P. and Birrer, S. and Spiniello, C. and Treu, T.},
  title = {{Strong Lensing by Galaxies}},
  journal = {Space Science Reviews},
  year = {2024},
  volume = {220},
  number = {8},
  eid = {87},
  pages = {87},
  doi = {10.1007/s11214-024-01105-x},
  archiveprefix = {arXiv},
  eprint = {2210.10790},
  primaryclass = {astro-ph.GA},
  adsurl = {https://ui.adsabs.harvard.edu/abs/2024SSRv..220...87S},
  url = {https://doi.org/10.1007/s11214-024-01105-x},
}

@inproceedings{saharia2022palette,
  title = {{Palette: Image-to-Image Diffusion Models}},
  author = {Saharia, Chitwan and Chan, William and Chang, Huiwen and Lee, Chris and Ho, Jonathan and Salimans, Tim and Fleet, David and Norouzi, Mohammad},
  booktitle = {ACM SIGGRAPH          Conference Proceedings},
  pages = {1--10},
  year = {2022},
  doi = {10.1145/3528233.3530757},
  url = {https://doi.org/10.1145/3528233.3530757},
}

@article{ivezic2019lsst,
  author = {Ivezić, Željko and Kahn, Steven M. and Tyson, J. Anthony and Abel, Bob and Acosta, Emily and Allsman, Robyn and Alonso, David and AlSayyad, Yusra and Anderson, Scott F. and Andrew, John and P. Angel, James Roger and Angeli, George Z. and Ansari, Reza and Antilogus, Pierre and Araujo, Constanza and Armstrong, Robert and Arndt, Kirk T. and Astier, Pierre and Aubourg, Éric and Auza, Nicole and Axelrod, Tim S. and Bard, Deborah J. and Barr, Jeff D. and Barrau, Aurelian and Bartlett, James G. and Bauer, Amanda E. and Bauman, Brian J. and Baumont, Sylvain and Bechtol, Ellen and Bechtol, Keith and Becker, Andrew C. and Becla, Jacek and Beldica, Cristina and Bellavia, Steve and Bianco, Federica B. and Biswas, Rahul and Blanc, Guillaume and Blazek, Jonathan and Blandford, Roger D. and Bloom, Josh S. and Bogart, Joanne and Bond, Tim W. and Booth, Michael T. and Borgland, Anders W. and Borne, Kirk and Bosch, James F. and Boutigny, Dominique and Brackett, Craig A. and Bradshaw, Andrew and Brandt, William Nielsen and Brown, Michael E. and Bullock, James S. and Burchat, Patricia and Burke, David L. and Cagnoli, Gianpietro and Calabrese, Daniel and Callahan, Shawn and Callen, Alice L. and Carlin, Jeffrey L. and Carlson, Erin L. and Chandrasekharan, Srinivasan and Charles-Emerson, Glenaver and Chesley, Steve and Cheu, Elliott C. and Chiang, Hsin-Fang and Chiang, James and Chirino, Carol and Chow, Derek and Ciardi, David R. and Claver, Charles F. and Cohen-Tanugi, Johann and Cockrum, Joseph J. and Coles, Rebecca and Connolly, Andrew J. and Cook, Kem H. and Cooray, Asantha and Covey, Kevin R. and Cribbs, Chris and Cui, Wei and Cutri, Roc and Daly, Philip N. and Daniel, Scott F. and Daruich, Felipe and Daubard, Guillaume and Daues, Greg and Dawson, William and Delgado, Francisco and Dellapenna, Alfred and Peyster, Robert de and Val-Borro, Miguel de and Digel, Seth W. and Doherty, Peter and Dubois, Richard and Dubois-Felsmann, Gregory P. and Durech, Josef and Economou, Frossie and Eifler, Tim and Eracleous, Michael and Emmons, Benjamin L. and Neto, Angelo Fausti and Ferguson, Henry and Figueroa, Enrique and Fisher-Levine, Merlin and Focke, Warren and Foss, Michael D. and Frank, James and Freemon, Michael D. and Gangler, Emmanuel and Gawiser, Eric and Geary, John C. and Gee, Perry and Geha, Marla and Gessner, Charles J. B. and Gibson, Robert R. and Gilmore, D. Kirk and Glanzman, Thomas and Glick, William and Goldina, Tatiana and Goldstein, Daniel A. and Goodenow, Iain and Graham, Melissa L. and Gressler, William J. and Gris, Philippe and Guy, Leanne P. and Guyonnet, Augustin and Haller, Gunther and Harris, Ron and Hascall, Patrick A. and Haupt, Justine and Hernandez, Fabio and Herrmann, Sven and Hileman, Edward and Hoblitt, Joshua and Hodgson, John A. and Hogan, Craig and Howard, James D. and Huang, Dajun and Huffer, Michael E. and Ingraham, Patrick and Innes, Walter R. and Jacoby, Suzanne H. and Jain, Bhuvnesh and Jammes, Fabrice and Jee, M. James and Jenness, Tim and Jernigan, Garrett and Jevremović, Darko and Johns, Kenneth and Johnson, Anthony S. and Johnson, Margaret W. G. and Jones, R. Lynne and Juramy-Gilles, Claire and Jurić, Mario and Kalirai, Jason S. and Kallivayalil, Nitya J. and Kalmbach, Bryce and Kantor, Jeffrey P. and Karst, Pierre and Kasliwal, Mansi M. and Kelly, Heather and Kessler, Richard and Kinnison, Veronica and Kirkby, David and Knox, Lloyd and Kotov, Ivan V. and Krabbendam, Victor L. and Krughoff, K. Simon and Kubánek, Petr and Kuczewski, John and Kulkarni, Shri and Ku, John and Kurita, Nadine R. and Lage, Craig S. and Lambert, Ron and Lange, Travis and Langton, J. Brian and Guillou, Laurent Le and Levine, Deborah and Liang, Ming and Lim, Kian-Tat and Lintott, Chris J. and Long, Kevin E. and Lopez, Margaux and Lotz, Paul J. and Lupton, Robert H. and Lust, Nate B. and MacArthur, Lauren A. and Mahabal, Ashish and Mandelbaum, Rachel and Markiewicz, Thomas W. and Marsh, Darren S. and Marshall, Philip J. and Marshall, Stuart and May, Morgan and McKercher, Robert and McQueen, Michelle and Meyers, Joshua and Migliore, Myriam and Miller, Michelle and Mills, David J. and Miraval, Connor and Moeyens, Joachim and Moolekamp, Fred E. and Monet, David G. and Moniez, Marc and Monkewitz, Serge and Montgomery, Christopher and Morrison, Christopher B. and Mueller, Fritz and Muller, Gary P. and Arancibia, Freddy Muñoz and Neill, Douglas R. and Newbry, Scott P. and Nief, Jean-Yves and Nomerotski, Andrei and Nordby, Martin and O’Connor, Paul and Oliver, John and Olivier, Scot S. and Olsen, Knut and O’Mullane, William and Ortiz, Sandra and Osier, Shawn and Owen, Russell E. and Pain, Reynald and Palecek, Paul E. and Parejko, John K. and Parsons, James B. and Pease, Nathan M. and Peterson, J. Matt and Peterson, John R. and Petravick, Donald L. and Petrick, M. E. Libby and Petry, Cathy E. and Pierfederici, Francesco and Pietrowicz, Stephen and Pike, Rob and Pinto, Philip A. and Plante, Raymond and Plate, Stephen and Plutchak, Joel P. and Price, Paul A. and Prouza, Michael and Radeka, Veljko and Rajagopal, Jayadev and Rasmussen, Andrew P. and Regnault, Nicolas and Reil, Kevin A. and Reiss, David J. and Reuter, Michael A. and Ridgway, Stephen T. and Riot, Vincent J. and Ritz, Steve and Robinson, Sean and Roby, William and Roodman, Aaron and Rosing, Wayne and Roucelle, Cecille and Rumore, Matthew R. and Russo, Stefano and Saha, Abhijit and Sassolas, Benoit and Schalk, Terry L. and Schellart, Pim and Schindler, Rafe H. and Schmidt, Samuel and Schneider, Donald P. and Schneider, Michael D. and Schoening, William and Schumacher, German and Schwamb, Megan E. and Sebag, Jacques and Selvy, Brian and Sembroski, Glenn H. and Seppala, Lynn G. and Serio, Andrew and Serrano, Eduardo and Shaw, Richard A. and Shipsey, Ian and Sick, Jonathan and Silvestri, Nicole and Slater, Colin T. and Smith, J. Allyn and Smith, R. Chris and Sobhani, Shahram and Soldahl, Christine and Storrie-Lombardi, Lisa and Stover, Edward and Strauss, Michael A. and Street, Rachel A. and Stubbs, Christopher W. and Sullivan, Ian S. and Sweeney, Donald and Swinbank, John D. and Szalay, Alexander and Takacs, Peter and Tether, Stephen A. and Thaler, Jon J. and Thayer, John Gregg and Thomas, Sandrine and Thornton, Adam J. and Thukral, Vaikunth and Tice, Jeffrey and Trilling, David E. and Turri, Max and Berg, Richard Van and Berk, Daniel Vanden and Vetter, Kurt and Virieux, Francoise and Vucina, Tomislav and Wahl, William and Walkowicz, Lucianne and Walsh, Brian and Walter, Christopher W. and Wang, Daniel L. and Wang, Shin-Yawn and Warner, Michael and Wiecha, Oliver and Willman, Beth and Winters, Scott E. and Wittman, David and Wolff, Sidney C. and Wood-Vasey, W. Michael and Wu, Xiuqin and Xin, Bo and Yoachim, Peter and Zhan, Hu},
  title = {{LSST: From Science Drivers to Reference Design and Anticipated Data Products}},
  journal = {The Astrophysical Journal},
  year = {2019},
  volume = {873},
  number = {2},
  eid = {111},
  pages = {111},
  doi = {10.3847/1538-4357/ab042c},
  url = {https://doi.org/10.3847/1538-4357/ab042c},
}

@article{mellier2025euclid,
  author = {{Euclid Collaboration} and Mellier, Y. and {Abdurro’uf} and Acevedo Barroso, J. A. and Achúcarro, A. and Adamek, J. and Adam, R. and Addison, G. E. and Aghanim, N. and Aguena, M. and Ajani, V. and Akrami, Y. and Al-Bahlawan, A. and Alavi, A. and Albuquerque, I. S. and Alestas, G. and Alguero, G. and Allaoui, A. and Allen, S. W. and Allevato, V. and Alonso-Tetilla, A. V. and Altieri, B. and Alvarez-Candal, A. and Alvi, S. and Amara, A. and Amendola, L. and Amiaux, J. and Andika, I. T. and Andreon, S. and Andrews, A. and Angora, G. and Angulo, R. E. and Annibali, F. and Anselmi, A. and Anselmi, S. and Arcari, S. and Archidiacono, M. and Aricò, G. and Arnaud, M. and Arnouts, S. and Asgari, M. and Asorey, J. and Atayde, L. and Atek, H. and Atrio-Barandela, F. and Aubert, M. and Aubourg, E. and Auphan, T. and Auricchio, N. and Aussel, B. and Aussel, H. and Avelino, P. P. and Avgoustidis, A. and Avila, S. and Awan, S. and Azzollini, R. and Baccigalupi, C. and Bachelet, E. and Bacon, D. and Baes, M. and Bagley, M. B. and Bahr-Kalus, B. and Balaguera-Antolinez, A. and Balbinot, E. and Balcells, M. and Baldi, M. and Baldry, I. and Balestra, A. and Ballardini, M. and Ballester, O. and Balogh, M. and Bañados, E. and Barbier, R. and Bardelli, S. and Baron, M. and Barreiro, T. and Barrena, R. and Barriere, J.-C. and Barros, B. J. and Barthelemy, A. and Bartolo, N. and Basset, A. and Battaglia, P. and Battisti, A. J. and Baugh, C. M. and Baumont, L. and Bazzanini, L. and Beaulieu, J.-P. and Beckmann, V. and Belikov, A. N. and Bel, J. and Bellagamba, F. and Bella, M. and Bellini, E. and Benabed, K. and Bender, R. and Benevento, G. and Bennett, C. L. and Benson, K. and Bergamini, P. and Bermejo-Climent, J. R. and Bernardeau, F. and Bertacca, D. and Berthe, M. and Berthier, J. and Bethermin, M. and Beutler, F. and Bevillon, C. and Bhargava, S. and Bhatawdekar, R. and Bianchi, D. and Bisigello, L. and Biviano, A. and Blake, R. P. and Blanchard, A. and Blazek, J. and Blot, L. and Bosco, A. and Bodendorf, C. and Boenke, T. and Böhringer, H. and Boldrini, P. and Bolzonella, M. and Bonchi, A. and Bonici, M. and Bonino, D. and Bonino, L. and Bonvin, C. and Bon, W. and Booth, J. T. and Borgani, S. and Borlaff, A. S. and Borsato, E. and Bosco, A. and Bose, B. and Botticella, M. T. and Boucaud, A. and Bouche, F. and Boucher, J. S. and Boutigny, D. and Bouvard, T. and Bouwens, R. and Bouy, H. and Bowler, R. A. A. and Bozza, V. and Bozzo, E. and Branchini, E. and Brando, G. and Brau-Nogue, S. and Brekke, P. and Bremer, M. N. and Brescia, M. and Breton, M.-A. and Brinchmann, J. and Brinckmann, T. and Brockley-Blatt, C. and Brodwin, M. and Brouard, L. and Brown, M. L. and Bruton, S. and Bucko, J. and Buddelmeijer, H. and Buenadicha, G. and Buitrago, F. and Burger, P. and Burigana, C. and Busillo, V. and Busonero, D. and Cabanac, R. and Cabayol-Garcia, L. and Cagliari, M. S. and Caillat, A. and Caillat, L. and Calabrese, M. and Calabro, A. and Calderone, G. and Calura, F. and Camacho Quevedo, B. and Camera, S. and Campos, L. and Cañas-Herrera, G. and Candini, G. P. and Cantiello, M. and Capobianco, V. and Cappellaro, E. and Cappelluti, N. and Cappi, A. and Caputi, K. I. and Cara, C. and Carbone, C. and Cardone, V. F. and Carella, E. and Carlberg, R. G. and Carle, M. and Carminati, L. and Caro, F. and Carrasco, J. M. and Carretero, J. and Carrilho, P. and Carron Duque, J. and Carry, B. and Carvalho, A. and Carvalho, C. S. and Casas, R. and Casas, S. and Casenove, P. and Casey, C. M. and Cassata, P. and Castander, F. J. and Castelao, D. and Castellano, M. and Castiblanco, L. and Castignani, G. and Castro, T. and Cavet, C. and Cavuoti, S. and Chabaud, P.-Y. and Chambers, K. C. and Charles, Y. and Charlot, S. and Chartab, N. and Chary, R. and Chaumeil, F. and Cho, H. and Chon, G. and Ciancetta, E. and Ciliegi, P. and Cimatti, A. and Cimino, M. and Cioni, M.-R. L. and Claydon, R. and Cleland, C. and Clément, B. and Clements, D. L. and Clerc, N. and Clesse, S. and Codis, S. and Cogato, F. and Colbert, J. and Cole, R. E. and Coles, P. and Collett, T. E. and Collins, R. S. and Colodro-Conde, C. and Colombo, C. and Combes, F. and Conforti, V. and Congedo, G. and Conseil, S. and Conselice, C. J. and Contarini, S. and Contini, T. and Conversi, L. and Cooray, A. R. and Copin, Y. and Corasaniti, P.-S. and Corcho-Caballero, P. and Corcione, L. and Cordes, O. and Corpace, O. and Correnti, M. and Costanzi, M. and Costille, A. and Courbin, F. and Courcoult Mifsud, L. and Courtois, H. M. and Cousinou, M.-C. and Covone, G. and Cowell, T. and Cragg, C. and Cresci, G. and Cristiani, S. and Crocce, M. and Cropper, M. and Crouzet, P. E. and Csizi, B. and Cuby, J.-G. and Cucchetti, E. and Cucciati, O. and Cuillandre, J.-C. and Cunha, P. A. C. and Cuozzo, V. and Daddi, E. and D’Addona, M. and Dafonte, C. and Dagoneau, N. and Dalessandro, E. and Dalton, G. B. and D’Amico, G. and Dannerbauer, H. and Danto, P. and Das, I. and Da Silva, A. and da Silva, R. and d’Assignies Doumerg, W. and Daste, G. and Davies, J. E. and Davini, S. and Dayal, P. and de Boer, T. and Decarli, R. and De Caro, B. and Degaudenzi, H. and Degni, G. and de Jong, J. T. A. and de la Bella, L. F. and de la Torre, S. and Delhaise, F. and Delley, D. and Delucchi, G. and De Lucia, G. and Denniston, J. and De Paolis, F. and De Petris, M. and Derosa, A. and Desai, S. and Desjacques, V. and Despali, G. and Desprez, G. and De Vicente-Albendea, J. and Deville, Y. and Dias, J. D. F. and Díaz-Sánchez, A. and Diaz, J. J. and Di Domizio, S. and Diego, J. M. and Di Ferdinando, D. and Di Giorgio, A. M. and Dimauro, P. and Dinis, J. and Dolag, K. and Dolding, C. and Dole, H. and Domínguez Sánchez, H. and Doré, O. and Dournac, F. and Douspis, M. and Dreihahn, H. and Droge, B. and Dryer, B. and Dubath, F. and Duc, P.-A. and Ducret, F. and Duffy, C. and Dufresne, F. and Duncan, C. A. J. and Dupac, X. and Duret, V. and Durrer, R. and Durret, F. and Dusini, S. and Ealet, A. and Eggemeier, A. and Eisenhardt, P. R. M. and Elbaz, D. and Elkhashab, M. Y. and Ellien, A. and Endicott, J. and Enia, A. and Erben, T. and Escartin Vigo, J. A. and Escoffier, S. and Escudero Sanz, I. and Essert, J. and Ettori, S. and Ezziati, M. and Fabbian, G. and Fabricius, M. and Fang, Y. and Farina, A. and Farina, M. and Farinelli, R. and Farrens, S. and Faustini, F. and Feltre, A. and Ferguson, A. M. N. and Ferrando, P. and Ferrari, A. G. and Ferré-Mateu, A. and Ferreira, P. G. and Ferreras, I. and Ferrero, I. and Ferriol, S. and Ferruit, P. and Filleul, D. and Finelli, F. and Finkelstein, S. L. and Finoguenov, A. and Fiorini, B. and Flentge, F. and Focardi, P. and Fonseca, J. and Fontana, A. and Fontanot, F. and Fornari, F. and Fosalba, P. and Fossati, M. and Fotopoulou, S. and Fouchez, D. and Fourmanoit, N. and Frailis, M. and Fraix-Burnet, D. and Franceschi, E. and Franco, A. and Franzetti, P. and Freihoefer, J. and Frenk, C. S. and Frittoli, G. and Frugier, P.-A. and Frusciante, N. and Fumagalli, A. and Fumagalli, M. and Fumana, M. and Fu, Y. and Gabarra, L. and Galeotta, S. and Galluccio, L. and Ganga, K. and Gao, H. and García-Bellido, J. and Garcia, K. and Gardner, J. P. and Garilli, B. and Gaspar-Venancio, L.-M. and Gasparetto, T. and Gautard, V. and Gavazzi, R. and Gaztanaga, E. and Genolet, L. and Genova Santos, R. and Gentile, F. and George, K. and Gerbino, M. and Ghaffari, Z. and Giacomini, F. and Gianotti, F. and Gibb, G. P. S. and Gillard, W. and Gillis, B. and Ginolfi, M. and Giocoli, C. and Girardi, M. and Giri, S. K. and Goh, L. W. K. and Gómez-Alvarez, P. and Gonzalez-Perez, V. and Gonzalez, A. H. and Gonzalez, E. J. and Gonzalez, J. C. and Gouyou Beauchamps, S. and Gozaliasl, G. and Gracia-Carpio, J. and Grandis, S. and Granett, B. R. and Granvik, M. and Grazian, A. and Gregorio, A. and Grenet, C. and Grillo, C. and Grupp, F. and Gruppioni, C. and Gruppuso, A. and Guerbuez, C. and Guerrini, S. and Guidi, M. and Guillard, P. and Gutierrez, C. M. and Guttridge, P. and Guzzo, L. and Gwyn, S. and Haapala, J. and Haase, J. and Haddow, C. R. and Hailey, M. and Hall, A. and Hall, D. and Hamaus, N. and Haridasu, B. S. and Harnois-Déraps, J. and Harper, C. and Hartley, W. G. and Hasinger, G. and Hassani, F. and Hatch, N. A. and Haugan, S. V. H. and Häußler, B. and Heavens, A. and Heisenberg, L. and Helmi, A. and Helou, G. and Hemmati, S. and Henares, K. and Herent, O. and Hernández-Monteagudo, C. and Heuberger, T. and Hewett, P. C. and Heydenreich, S. and Hildebrandt, H. and Hirschmann, M. and Hjorth, J. and Hoar, J. and Hoekstra, H. and Holland, A. D. and Holliman, M. S. and Holmes, W. and Hook, I. and Horeau, B. and Hormuth, F. and Hornstrup, A. and Hosseini, S. and Hu, D. and Hudelot, P. and Hudson, M. J. and Huertas-Company, M. and Huff, E. M. and Hughes, A. C. N. and Humphrey, A. and Hunt, L. K. and Huynh, D. D. and Ibata, R. and Ichikawa, K. and Iglesias-Groth, S. and Ilbert, O. and Ilić, S. and Ingoglia, L. and Iodice, E. and Israel, H. and Israelsson, U. E. and Izzo, L. and Jablonka, P. and Jackson, N. and Jacobson, J. and Jafariyazani, M. and Jahnke, K. and Jain, B. and Jansen, H. and Jarvis, M. J. and Jasche, J. and Jauzac, M. and Jeffrey, N. and Jhabvala, M. and Jimenez-Teja, Y. and Jimenez Muñoz, A. and Joachimi, B. and Johansson, P. H. and Joudaki, S. and Jullo, E. and Kajava, J. J. E. and Kang, Y. and Kannawadi, A. and Kansal, V. and Karagiannis, D. and Kärcher, M. and Kashlinsky, A. and Kazandjian, M. V. and Keck, F. and Keihänen, E. and Kerins, E. and Kermiche, S. and Khalil, A. and Kiessling, A. and Kiiveri, K. and Kilbinger, M. and Kim, J. and King, R. and Kirkpatrick, C. C. and Kitching, T. and Kluge, M. and Knabenhans, M. and Knapen, J. H. and Knebe, A. and Kneib, J.-P. and Kohley, R. and Koopmans, L. V. E. and Koskinen, H. and Koulouridis, E. and Kou, R. and Kovács, A. and Kovačić, I. and Kowalczyk, A. and Koyama, K. and Kraljic, K. and Krause, O. and Kruk, S. and Kubik, B. and Kuchner, U. and Kuijken, K. and Kümmel, M. and Kunz, M. and Kurki-Suonio, H. and Lacasa, F. and Lacey, C. G. and La Franca, F. and Lagarde, N. and Lahav, O. and Laigle, C. and La Marca, A. and La Marle, O. and Lamine, B. and Lam, M. C. and Lançon, A. and Landt, H. and Langer, M. and Lapi, A. and Larcheveque, C. and Larsen, S. S. and Lattanzi, M. and Laudisio, F. and Laugier, D. and Laureijs, R. and Laurent, V. and Lavaux, G. and Lawrenson, A. and Lazanu, A. and Lazeyras, T. and Le Boulc’h, Q. and Le Brun, A. M. C. and Le Brun, V. and Leclercq, F. and Lee, S. and Le Graet, J. and Legrand, L. and Leirvik, K. N. and Le Jeune, M. and Lembo, M. and Le Mignant, D. and Lepinzan, M. D. and Lepori, F. and Le Reun, A. and Leroy, G. and Lesci, G. F. and Lesgourgues, J. and Leuzzi, L. and Levi, M. E. and Liaudat, T. I. and Libet, G. and Liebing, P. and Ligori, S. and Lilje, P. B. and Lin, C.-C. and Linde, D. and Linder, E. and Lindholm, V. and Linke, L. and Li, S.-S. and Liu, S. J. and Lloro, I. and Lobo, F. S. N. and Lodieu, N. and Lombardi, M. and Lombriser, L. and Lonare, P. and Longo, G. and López-Caniego, M. and Lopez Lopez, X. and Lorenzo Alvarez, J. and Loureiro, A. and Loveday, J. and Lusso, E. and Macias-Perez, J. and Maciaszek, T. and Maggio, G. and Magliocchetti, M. and Magnard, F. and Magnier, E. A. and Magro, A. and Mahler, G. and Mainetti, G. and Maino, D. and Maiorano, E. and Maiorano, E. and Malavasi, N. and Mamon, G. A. and Mancini, C. and Mandelbaum, R. and Manera, M. and Manjón-García, A. and Mannucci, F. and Mansutti, O. and Manteiga Outeiro, M. and Maoli, R. and Maraston, C. and Marcin, S. and Marcos-Arenal, P. and Margalef-Bentabol, B. and Marggraf, O. and Marinucci, D. and Marinucci, M. and Markovic, K. and Marleau, F. R. and Marpaud, J. and Martignac, J. and Martín-Fleitas, J. and Martin-Moruno, P. and Martin, E. L. and Martinelli, M. and Martinet, N. and Martin, H. and Martins, C. J. A. P. and Marulli, F. and Massari, D. and Massey, R. and Masters, D. C. and Matarrese, S. and Matsuoka, Y. and Matthew, S. and Maughan, B. J. and Mauri, N. and Maurin, L. and Maurogordato, S. and McCarthy, K. and McConnachie, A. W. and McCracken, H. J. and McDonald, I. and McEwen, J. D. and McPartland, C. J. R. and Medinaceli, E. and Mehta, V. and Mei, S. and Melchior, M. and Melin, J.-B. and Ménard, B. and Mendes, J. and Mendez-Abreu, J. and Meneghetti, M. and Mercurio, A. and Merlin, E. and Metcalf, R. B. and Meylan, G. and Migliaccio, M. and Mignoli, M. and Miller, L. and Miluzio, M. and Milvang-Jensen, B. and Mimoso, J. P. and Miquel, R. and Miyatake, H. and Mobasher, B. and Mohr, J. J. and Monaco, P. and Monguió, M. and Montoro, A. and Mora, A. and Moradinezhad Dizgah, A. and Moresco, M. and Moretti, C. and Morgante, G. and Morisset, N. and Moriya, T. J. and Morris, P. W. and Mortlock, D. J. and Moscardini, L. and Mota, D. F. and Mottet, S. and Moustakas, L. A. and Moutard, T. and Müller, T. and Munari, E. and Murphree, G. and Murray, C. and Murray, N. and Musi, P. and Nadathur, S. and Nagam, B. C. and Nagao, T. and Naidoo, K. and Nakajima, R. and Nally, C. and Natoli, P. and Navarro-Alsina, A. and Navarro Girones, D. and Neissner, C. and Nersesian, A. and Nesseris, S. and Nguyen-Kim, H. N. and Nicastro, L. and Nichol, R. C. and Nielbock, M. and Niemi, S.-M. and Nieto, S. and Nilsson, K. and Noller, J. and Norberg, P. and Nouri-Zonoz, A. and Ntelis, P. and Nucita, A. A. and Nugent, P. and Nunes, N. J. and Nutma, T. and Ocampo, I. and Odier, J. and Oesch, P. A. and Oguri, M. and Magalhaes Oliveira, D. and Onoue, M. and Oosterbroek, T. and Oppizzi, F. and Ordenovic, C. and Osato, K. and Pacaud, F. and Pace, F. and Padilla, C. and Paech, K. and Pagano, L. and Page, M. J. and Palazzi, E. and Paltani, S. and Pamuk, S. and Pandolfi, S. and Paoletti, D. and Paolillo, M. and Papaderos, P. and Pardede, K. and Parimbelli, G. and Parmar, A. and Partmann, C. and Pasian, F. and Passalacqua, F. and Paterson, K. and Patrizii, L. and Pattison, C. and Paulino-Afonso, A. and Paviot, R. and Peacock, J. A. and Pearce, F. R. and Pedersen, K. and Peel, A. and Peletier, R. F. and Pellejero Ibanez, M. and Pello, R. and Penny, M. T. and Percival, W. J. and Perez-Garrido, A. and Perotto, L. and Pettorino, V. and Pezzotta, A. and Pezzuto, S. and Philippon, A. and Pierre, M. and Piersanti, O. and Pietroni, M. and Piga, L. and Pilo, L. and Pires, S. and Pisani, A. and Pizzella, A. and Pizzuti, L. and Plana, C. and Polenta, G. and Pollack, J. E. and Poncet, M. and Pöntinen, M. and Pool, P. and Popa, L. A. and Popa, V. and Popp, J. and Porciani, C. and Porth, L. and Potter, D. and Poulain, M. and Pourtsidou, A. and Pozzetti, L. and Prandoni, I. and Pratt, G. W. and Prezelus, S. and Prieto, E. and Pugno, A. and Quai, S. and Quilley, L. and Racca, G. D. and Raccanelli, A. and Rácz, G. and Radinović, S. and Radovich, M. and Ragagnin, A. and Ragnit, U. and Raison, F. and Ramos-Chernenko, N. and Ranc, C. and Rasera, Y. and Raylet, N. and Rebolo, R. and Refregier, A. and Reimberg, P. and Reiprich, T. H. and Renk, F. and Renzi, A. and Retre, J. and Revaz, Y. and Reylé, C. and Reynolds, L. and Rhodes, J. and Ricci, F. and Ricci, M. and Riccio, G. and Ricken, S. O. and Rissanen, S. and Risso, I. and Rix, H.-W. and Robin, A. C. and Rocca-Volmerange, B. and Rocci, P.-F. and Rodenhuis, M. and Rodighiero, G. and Rodriguez Monroy, M. and Rollins, R. P. and Romanello, M. and Roman, J. and Romelli, E. and Romero-Gomez, M. and Roncarelli, M. and Rosati, P. and Rosset, C. and Rossetti, E. and Roster, W. and Rottgering, H. J. A. and Rozas-Fernández, A. and Ruane, K. and Rubino-Martin, J. A. and Rudolph, A. and Ruppin, F. and Rusholme, B. and Sacquegna, S. and Sáez-Casares, I. and Saga, S. and Saglia, R. and Sahlén, M. and Saifollahi, T. and Sakr, Z. and Salvalaggio, J. and Salvaterra, R. and Salvati, L. and Salvato, M. and Salvignol, J.-C. and Sánchez, A. G. and Sanchez, E. and Sanders, D. B. and Sapone, D. and Saponara, M. and Sarpa, E. and Sarron, F. and Sartori, S. and Sartoris, B. and Sassolas, B. and Sauniere, L. and Sauvage, M. and Sawicki, M. and Scaramella, R. and Scarlata, C. and Scharré, L. and Schaye, J. and Schewtschenko, J. A. and Schindler, J.-T. and Schinnerer, E. and Schirmer, M. and Schmidt, F. and Schmidt, F. and Schmidt, M. and Schneider, A. and Schneider, M. and Schneider, P. and Schöneberg, N. and Schrabback, T. and Schultheis, M. and Schulz, S. and Schuster, N. and Schwartz, J. and Sciotti, D. and Scodeggio, M. and Scognamiglio, D. and Scott, D. and Scottez, V. and Secroun, A. and Sefusatti, E. and Seidel, G. and Seiffert, M. and Sellentin, E. and Selwood, M. and Semboloni, E. and Sereno, M. and Serjeant, S. and Serrano, S. and Setnikar, G. and Shankar, F. and Sharples, R. M. and Short, A. and Shulevski, A. and Shuntov, M. and Sias, M. and Sikkema, G. and Silvestri, A. and Simon, P. and Sirignano, C. and Sirri, G. and Skottfelt, J. and Slezak, E. and Sluse, D. and Smith, G. P. and Smith, L. C. and Smith, R. E. and Smit, S. J. A. and Soldano, F. and Solheim, B. G. B. and Sorce, J. G. and Sorrenti, F. and Soubrie, E. and Spinoglio, L. and Spurio Mancini, A. and Stadel, J. and Stagnaro, L. and Stanco, L. and Stanford, S. A. and Starck, J.-L. and Stassi, P. and Steinwagner, J. and Stern, D. and Stone, C. and Strada, P. and Strafella, F. and Stramaccioni, D. and Surace, C. and Sureau, F. and Suyu, S. H. and Swindells, I. and Szafraniec, M. and Szapudi, I. and Taamoli, S. and Talia, M. and Tallada-Crespí, P. and Tanidis, K. and Tao, C. and Tarrío, P. and Tavagnacco, D. and Taylor, A. N. and Taylor, J. E. and Taylor, P. L. and Teixeira, E. M. and Tenti, M. and Teodoro Idiago, P. and Teplitz, H. I. and Tereno, I. and Tessore, N. and Testa, V. and Testera, G. and Tewes, M. and Teyssier, R. and Theret, N. and Thizy, C. and Thomas, P. D. and Toba, Y. and Toft, S. and Toledo-Moreo, R. and Tolstoy, E. and Tommasi, E. and Torbaniuk, O. and Torradeflot, F. and Tortora, C. and Tosi, S. and Tosti, S. and Trifoglio, M. and Troja, A. and Trombetti, T. and Tronconi, A. and Tsedrik, M. and Tsyganov, A. and Tucci, M. and Tutusaus, I. and Uhlemann, C. and Ulivi, L. and Urbano, M. and Vacher, L. and Vaillon, L. and Valageas, P. and Valdes, I. and Valentijn, E. A. and Valenziano, L. and Valieri, C. and Valiviita, J. and Van den Broeck, M. and Vassallo, T. and Vavrek, R. and Vega-Ferrero, J. and Venemans, B. and Venhola, A. and Ventura, S. and Verdoes Kleijn, G. and Vergani, D. and Verma, A. and Vernizzi, F. and Veropalumbo, A. and Verza, G. and Vescovi, C. and Vibert, D. and Viel, M. and Vielzeuf, P. and Viglione, C. and Viitanen, A. and Villaescusa-Navarro, F. and Vinciguerra, S. and Visticot, F. and Voggel, K. and von Wietersheim-Kramsta, M. and Vriend, W. J. and Wachter, S. and Walmsley, M. and Walth, G. and Walton, D. M. and Walton, N. A. and Wander, M. and Wang, L. and Wang, Y. and Weaver, J. R. and Weller, J. and Wetzstein, M. and Whalen, D. J. and Whittam, I. H. and Widmer, A. and Wiesmann, M. and Wilde, J. and Williams, O. R. and Winther, H.-A. and Wittje, A. and Wong, J. H. W. and Wright, A. H. and Yankelevich, V. and Yeung, H. W. and Yoon, M. and Youles, S. and Yung, L. Y. A. and Zacchei, A. and Zalesky, L. and Zamorani, G. and Zamorano Vitorelli, A. and Zanoni Marc, M. and Zennaro, M. and Zerbi, F. M. and Zinchenko, I. A. and Zoubian, J. and Zucca, E. and Zumalacarregui, M.},
  title = {{Euclid: I. Overview of the Euclid mission}},
  journal = {Astronomy \& Astrophysics},
  year = {2025},
  volume = {697},
  eid = {A1},
  pages = {A1},
  doi = {10.1051/0004-6361/202450810},
  url = {https://doi.org/10.1051/0004-6361/202450810},
}

@article{khadka2026slsim,
  author = {Khadka, Narayan and Birrer, Simon and Best, Henry and Sharma, Paras and Abe, Katsuya T. and Tang, Xianzhe and Mistick, Carly and Urcelay, Felipe and Sonmez, Emrecan M. and Arendse, Nikki and Erickson, Sydney and Hjortlund, Jacob O. and Holloway, Phil and Huang, Alan and Karthik, Rahul and Lamontagne, Mia and Negi, Vibhore and Pierel, Justin R. and Sanchez, Bruno and Ece Saricaoglu, Aysu and Shajib, Anowar and Shao, Yixuan and Venkatraman, Padma and Wedig, Bryce and Agrawal, Aadya and Anguita, Timo and Bessa, Pedro and Bom, Clecio R. and Castillo, Sofia and Collett, Thomas and Daylan, Tansu and Dillmann, Steven and Grespan, Margherita and Hayes, Erin E. and Joseph, Remy and Kessler, Richard and Li, Tian and Marshall, Phil and More, Anupreeta and Motta, Veronica and Narayan, Gautham and O'Dowd, Matt and Oguri, Masamune and Smith, Graham and Verma, Aprajita and Vernardos, Giorgos and {the Strong Lensing Science Collaboration} and {the LSST Dark Energy Science Collaboration}},
  title = {{SLSim: a strong lensing population simulation package}},
  journal = {arXiv e-prints},
  year = {2026},
  pages = {arXiv:2603.17138},
  doi = {10.48550/arXiv.2603.17138},
  archivePrefix = {arXiv},
  eprint = {2603.17138},
  primaryClass = {astro-ph.CO},
  url = {https://arxiv.org/abs/2603.17138},
}

@misc{speclite_zenodo_2024,
  author = {David Kirkby and Thomas Robitaille and Benjamin Alan Weaver and {Moustakas} and Erik Tollerud and Michael Droettboom and Brigitta Sipőcz and E. M. Bray and Larry Bradley and Andy Park and J. Michael Burgess and Marcelo Alvarez and Stephen Bailey and Sergey Koposov and Dustin Lang and Matt Craig and Christoph Deil and P. L. Lim and Kyle Barbary and Adam Ginsburg and Wolfgang Kerzendorf and {dcunning11235} and Hélion du Mas des Bourboux and Steve Crawford and Víctor Zabalza and Javier Sanchez and Hans Moritz Günther},
  title = {{desihub/speclite: Bug fix release: General clean-up prior to refactoring package infrastructure}},
  year = {2024},
  publisher = {Zenodo},
  version = {v0.20},
  doi = {10.5281/zenodo.13225530},
  url = {https://doi.org/10.5281/zenodo.13225530},
  howpublished = {Zenodo software release},
  note = {Version v0.20},
}

@article{Amara2021,
  doi = {10.21105/joss.03056},
  year = {2021},
  publisher = {The Open Journal},
  volume = {6},
  number = {65},
  pages = {3056},
  author = {Amara, Adam and de la Bella, Lucia and Birrer, Simon and Bridle, Sarah and Cordero, Juan and Favole, Ginevra and Harrison, Ian and Harry, Ian and Hartley, William and Krawczyk, Coleman and Lundgren, Andrew and Nord, Brian and Nuttall, Laura and Rollins, Richard and Sudek, Philipp and Tam, Sut-Ieng and Tessore, Nicolas and Tolley, Arthur and Umetsu, Keiichi and Williamson, Andrew and Wolz, Laura},
  title = {{SkyPy: A package for modelling the Universe}},
  journal = {Journal of Open Source Software},
  url = {https://doi.org/10.21105/joss.03056},
}

@article{Gneiting01032007,
  author = {Gneiting, Tilmann and Raftery, Adrian E},
  title = {{Strictly Proper Scoring Rules, Prediction, and Estimation}},
  journal = {Journal of the American Statistical Association},
  volume = {102},
  number = {477},
  pages = {359--378},
  year = {2007},
  publisher = {Taylor \& Francis},
  doi = {10.1198/016214506000001437},
  url = {https://doi.org/10.1198/016214506000001437},
}

@article{roman,
  author = {Mosby, Gregory and Rauscher, Bernard and Bennett, Chris and Cheng, Edward S. and Cheung, Stephanie A. and Cillis, Analia N. and Content, David A. and Cottingham, David A. and Foltz, Roger D. and Gygax, John D. and Hill, Robert J. and Kruk, Jeffrey W. and Mah, Jon S. and Meier, Lane A. and Merchant, Chris A. and Miko, Laddawan R. and Piquette, Eric C. and Waczynski, Augustyn and Wen, Yiting},
  title = {{Properties and characteristics of the Nancy Grace Roman Space Telescope H4RG-10 detectors}},
  journal = {Journal of Astronomical Telescopes, Instruments, and Systems},
  year = {2020},
  volume = {6},
  eid = {046001},
  pages = {046001},
  doi = {10.1117/1.jatis.6.4.046001},
  url = {https://doi.org/10.1117/1.jatis.6.4.046001},
  number = {4},
}

@article{simon,
  year = {2019},
  publisher = {},
  volume = {2019},
  number = {02},
  pages = {056},
  author = {Ade, Peter and Aguirre, James and Ahmed, Zeeshan and Aiola, Simone and Ali, Aamir and Alonso, David and Alvarez, Marcelo A. and Arnold, Kam and Ashton, Peter and Austermann, Jason and Awan, Humna and Baccigalupi, Carlo and Baildon, Taylor and Barron, Darcy and Battaglia, Nick and Battye, Richard and Baxter, Eric and Bazarko, Andrew and Beall, James A. and Bean, Rachel and Beck, Dominic and Beckman, Shawn and Beringue, Benjamin and Bianchini, Federico and Boada, Steven and Boettger, David and Bond, J. Richard and Borrill, Julian and Brown, Michael L. and Bruno, Sarah Marie and Bryan, Sean and Calabrese, Erminia and Calafut, Victoria and Calisse, Paolo and Carron, Julien and Challinor, Anthony and Chesmore, Grace and Chinone, Yuji and Chluba, Jens and Cho, Hsiao-Mei Sherry and Choi, Steve and Coppi, Gabriele and Cothard, Nicholas F. and Coughlin, Kevin and Crichton, Devin and Crowley, Kevin D. and Crowley, Kevin T. and Cukierman, Ari and D'Ewart, John M. and Dünner, Rolando and de Haan, Tijmen and Devlin, Mark and Dicker, Simon and Didier, Joy and Dobbs, Matt and Dober, Bradley and Duell, Cody J. and Duff, Shannon and Duivenvoorden, Adri and Dunkley, Jo and Dusatko, John and Errard, Josquin and Fabbian, Giulio and Feeney, Stephen and Ferraro, Simone and Fluxà, Pedro and Freese, Katherine and Frisch, Josef C. and Frolov, Andrei and Fuller, George and Fuzia, Brittany and Galitzki, Nicholas and Gallardo, Patricio A. and Ghersi, Jose Tomas Galvez and Gao, Jiansong and Gawiser, Eric and Gerbino, Martina and Gluscevic, Vera and Goeckner-Wald, Neil and Golec, Joseph and Gordon, Sam and Gralla, Megan and Green, Daniel and Grigorian, Arpi and Groh, John and Groppi, Chris and Guan, Yilun and Gudmundsson, Jon E. and Han, Dongwon and Hargrave, Peter and Hasegawa, Masaya and Hasselfield, Matthew and Hattori, Makoto and Haynes, Victor and Hazumi, Masashi and He, Yizhou and Healy, Erin and Henderson, Shawn W. and Hervias-Caimapo, Carlos and Hill, Charles A. and Hill, J. Colin and Hilton, Gene and Hilton, Matt and Hincks, Adam D. and Hinshaw, Gary and Hložek, Renée and Ho, Shirley and Ho, Shuay-Pwu Patty and Howe, Logan and Huang, Zhiqi and Hubmayr, Johannes and Huffenberger, Kevin and Hughes, John P. and Ijjas, Anna and Ikape, Margaret and Irwin, Kent and Jaffe, Andrew H. and Jain, Bhuvnesh and Jeong, Oliver and Kaneko, Daisuke and Karpel, Ethan D. and Katayama, Nobuhiko and Keating, Brian and Kernasovskiy, Sarah S. and Keskitalo, Reijo and Kisner, Theodore and Kiuchi, Kenji and Klein, Jeff and Knowles, Kenda and Koopman, Brian and Kosowsky, Arthur and Krachmalnicoff, Nicoletta and Kuenstner, Stephen E. and Kuo, Chao-Lin and Kusaka, Akito and Lashner, Jacob and Lee, Adrian and Lee, Eunseong and Leon, David and Leung, Jason S.-Y. and Lewis, Antony and Li, Yaqiong and Li, Zack and Limon, Michele and Linder, Eric and Lopez-Caraballo, Carlos and Louis, Thibaut and Lowry, Lindsay and Lungu, Marius and Madhavacheril, Mathew and Mak, Daisy and Maldonado, Felipe and Mani, Hamdi and Mates, Ben and Matsuda, Frederick and Maurin, Loïc and Mauskopf, Phil and May, Andrew and McCallum, Nialh and McKenney, Chris and McMahon, Jeff and Meerburg, P. Daniel and Meyers, Joel and Miller, Amber and Mirmelstein, Mark and Moodley, Kavilan and Munchmeyer, Moritz and Munson, Charles and Naess, Sigurd and Nati, Federico and Navaroli, Martin and Newburgh, Laura and Nguyen, Ho Nam and Niemack, Michael and Nishino, Haruki and Orlowski-Scherer, John and Page, Lyman and Partridge, Bruce and Peloton, Julien and Perrotta, Francesca and Piccirillo, Lucio and Pisano, Giampaolo and Poletti, Davide and Puddu, Roberto and Puglisi, Giuseppe and Raum, Chris and Reichardt, Christian L. and Remazeilles, Mathieu and Rephaeli, Yoel and Riechers, Dominik and Rojas, Felipe and Roy, Anirban and Sadeh, Sharon and Sakurai, Yuki and Salatino, Maria and Rao, Mayuri Sathyanarayana and Schaan, Emmanuel and Schmittfull, Marcel and Sehgal, Neelima and Seibert, Joseph and Seljak, Uros and Sherwin, Blake and Shimon, Meir and Sierra, Carlos and Sievers, Jonathan and Sikhosana, Precious and Silva-Feaver, Maximiliano and Simon, Sara M. and Sinclair, Adrian and Siritanasak, Praween and Smith, Kendrick and Smith, Stephen R. and Spergel, David and Staggs, Suzanne T. and Stein, George and Stevens, Jason R. and Stompor, Radek and Suzuki, Aritoki and Tajima, Osamu and Takakura, Satoru and Teply, Grant and Thomas, Daniel B. and Thorne, Ben and Thornton, Robert and Trac, Hy and Tsai, Calvin and Tucker, Carole and Ullom, Joel and Vagnozzi, Sunny and Engelen, Alexander van and Lanen, Jeff Van and Winkle, Daniel D. Van and Vavagiakis, Eve M. and Vergès, Clara and Vissers, Michael and Wagoner, Kasey and Walker, Samantha and Ward, Jon and Westbrook, Ben and Whitehorn, Nathan and Williams, Jason and Williams, Joel and Wollack, Edward J. and Xu, Zhilei and Yu, Byeonghee and Yu, Cyndia and Zago, Fernando and Zhang, Hezi and Zhu, Ningfeng and {The Simons Observatory collaboration}},
  title = {{The Simons Observatory: science goals and forecasts}},
  journal = {Journal of Cosmology and Astroparticle Physics},
  doi = {10.1088/1475-7516/2019/02/056},
  url = {https://doi.org/10.1088/1475-7516/2019/02/056},
}

@misc{spherex,
  title = {{Cosmology with the SPHEREX All-Sky Spectral Survey}},
  author = {Doré, Olivier and Bock, Jamie and Ashby, Matthew and Capak, Peter and Cooray, Asantha and de Putter, Roland and Eifler, Tim and Flagey, Nicolas and Gong, Yan and Habib, Salman and Heitmann, Katrin and Hirata, Chris and Jeong, Woong-Seob and Katti, Raj and Korngut, Phil and Krause, Elisabeth and Lee, Dae-Hee and Masters, Daniel and Mauskopf, Phil and Melnick, Gary and Mennesson, Bertrand and Nguyen, Hien and Öberg, Karin and Pullen, Anthony and Raccanelli, Alvise and Smith, Roger and Song, Yong-Seon and Tolls, Volker and Unwin, Steve and Venumadhav, Tejaswi and Viero, Marco and Werner, Mike and Zemcov, Mike},
  year = {2015},
  eprint = {1412.4872v3},
  archiveprefix = {arXiv},
  primaryclass = {astro-ph.CO},
  url = {https://arxiv.org/abs/1412.4872v3},
  doi = {10.48550/arXiv.1412.4872},
  howpublished = {arXiv preprint arXiv:1412.4872v3},
}

@article{collett2015population,
  title = {{THE POPULATION OF GALAXY–GALAXY STRONG LENSES IN FORTHCOMING OPTICAL IMAGING SURVEYS}},
  author = {Collett, Thomas E.},
  journal = {The Astrophysical Journal},
  volume = {811},
  number = {1},
  pages = {20},
  year = {2015},
  url = {https://doi.org/10.1088/0004-637x/811/1/20},
  doi = {10.1088/0004-637x/811/1/20},
}

@article{oguri2010statistics,
  title = {{Gravitationally lensed quasars and supernovae in future wide-field optical imaging surveys}},
  author = {Oguri, Masamune and Marshall, Philip J.},
  journal = {Monthly Notices of the Royal Astronomical Society},
  volume = {405},
  number = {4},
  pages = {2579--2593},
  year = {2010},
  url = {https://doi.org/10.1111/j.1365-2966.2010.16639.x},
  doi = {10.1111/j.1365-2966.2010.16639.x},
}

@article{lanusse2018cmu,
  title = {{CMU DeepLens: deep learning for automatic image-based galaxy–galaxy strong lens finding}},
  volume = {473},
  issn = {1365-2966},
  url = {https://doi.org/10.1093/mnras/stx1665},
  doi = {10.1093/mnras/stx1665},
  number = {3},
  journal = {Monthly Notices of the Royal Astronomical Society},
  publisher = {Oxford University Press (OUP)},
  author = {Lanusse, François and Ma, Quanbin and Li, Nan and Collett, Thomas E. and Li, Chun-Liang and Ravanbakhsh, Siamak and Mandelbaum, Rachel and Póczos, Barnabás},
  year = {2018},
  pages = {3895--3906},
}

@article{jacobs2019finding,
  title = {{Finding high-redshift strong lenses in DES using convolutional neural networks}},
  volume = {484},
  issn = {1365-2966},
  url = {https://doi.org/10.1093/mnras/stz272},
  doi = {10.1093/mnras/stz272},
  number = {4},
  journal = {Monthly Notices of the Royal Astronomical Society},
  publisher = {Oxford University Press (OUP)},
  author = {Jacobs, C and Collett, T and Glazebrook, K and McCarthy, C and Qin, A K and Abbott, T M C and Abdalla, F B and Annis, J and Avila, S and Bechtol, K and Bertin, E and Brooks, D and Buckley-Geer, E and Burke, D L and Carnero Rosell, A and Carrasco Kind, M and Carretero, J and da Costa, L N and Davis, C and De Vicente, J and Desai, S and Diehl, H T and Doel, P and Eifler, T F and Flaugher, B and Frieman, J and García-Bellido, J and Gaztanaga, E and Gerdes, D W and Goldstein, D A and Gruen, D and Gruendl, R A and Gschwend, J and Gutierrez, G and Hartley, W G and Hollowood, D L and Honscheid, K and Hoyle, B and James, D J and Kuehn, K and Kuropatkin, N and Lahav, O and Li, T S and Lima, M and Lin, H and Maia, M A G and Martini, P and Miller, C J and Miquel, R and Nord, B and Plazas, A A and Sanchez, E and Scarpine, V and Schubnell, M and Serrano, S and Sevilla-Noarbe, I and Smith, M and Soares-Santos, M and Sobreira, F and Suchyta, E and Swanson, M E C and Tarle, G and Vikram, V and Walker, A R and Zhang, Y and Zuntz, J and {DES Collaboration}},
  year = {2019},
  pages = {5330--5349},
}

@article{treu2010strong,
  title = {{Strong Lensing by Galaxies}},
  volume = {48},
  issn = {1545-4282},
  url = {https://doi.org/10.1146/annurev-astro-081309-130924},
  doi = {10.1146/annurev-astro-081309-130924},
  number = {1},
  journal = {Annual Review of Astronomy and Astrophysics},
  publisher = {Annual Reviews},
  author = {Treu, Tommaso},
  year = {2010},
  pages = {87--125},
}

@misc{kochanek2006strong,
  title = {{The Saas Fee Lectures on Strong Gravitational Lensing}},
  author = {Kochanek, C. S.},
  year = {2004},
  eprint = {astro-ph/0407232v1},
  archiveprefix = {arXiv},
  primaryclass = {astro-ph},
  url = {https://arxiv.org/abs/astro-ph/0407232v1},
  doi = {10.48550/arXiv.astro-ph/0407232},
  howpublished = {arXiv preprint arXiv:astro-ph/0407232v1},
}

@article{menze2015brats,
  author = {Menze, Bjoern H. and Jakab, Andras and Bauer, Stefan and Kalpathy-Cramer, Jayashree and Farahani, Keyvan and Kirby, Justin and Burren, Yuliya and Porz, Nicole and Slotboom, Johannes and Wiest, Roland and Lanczi, Levente and Gerstner, Elizabeth and Weber, Marc-Andre and Arbel, Tal and Avants, Brian B. and Ayache, Nicholas and Buendia, Patricia and Collins, D. Louis and Cordier, Nicolas and Corso, Jason J. and Criminisi, Antonio and Das, Tilak and Delingette, Herve and Demiralp, Cagatay and Durst, Christopher R. and Dojat, Michel and Doyle, Senan and Festa, Joana and Forbes, Florence and Geremia, Ezequiel and Glocker, Ben and Golland, Polina and Guo, Xiaotao and Hamamci, Andac and Iftekharuddin, Khan M. and Jena, Raj and John, Nigel M. and Konukoglu, Ender and Lashkari, Danial and Mariz, Jose Antonio and Meier, Raphael and Pereira, Sergio and Precup, Doina and Price, Stephen J. and Raviv, Tammy Riklin and Reza, Syed M. S. and Ryan, Michael and Sarikaya, Duygu and Schwartz, Lawrence and Shin, Hoo-Chang and Shotton, Jamie and Silva, Carlos A. and Sousa, Nuno and Subbanna, Nagesh K. and Szekely, Gabor and Taylor, Thomas J. and Thomas, Owen M. and Tustison, Nicholas J. and Unal, Gozde and Vasseur, Flor and Wintermark, Max and Ye, Dong Hye and Zhao, Liang and Zhao, Binsheng and Zikic, Darko and Prastawa, Marcel and Reyes, Mauricio and Van Leemput, Koen},
  journal = {IEEE Transactions on Medical Imaging},
  title = {{The Multimodal Brain Tumor Image Segmentation Benchmark (BRATS)}},
  year = {2015},
  volume = {34},
  number = {10},
  pages = {1993--2024},
  doi = {10.1109/tmi.2014.2377694},
  url = {https://doi.org/10.1109/tmi.2014.2377694},
}

@misc{bakas2018identifying,
  title = {{Identifying the Best Machine Learning Algorithms for Brain Tumor Segmentation, Progression Assessment, and Overall Survival Prediction in the BRATS Challenge}},
  author = {Bakas, Spyridon and Reyes, Mauricio and Jakab, Andras and Bauer, Stefan and Rempfler, Markus and Crimi, Alessandro and Shinohara, Russell Takeshi and Berger, Christoph and Ha, Sung Min and Rozycki, Martin and Prastawa, Marcel and Alberts, Esther and Lipkova, Jana and Freymann, John and Kirby, Justin and Bilello, Michel and Fathallah-Shaykh, Hassan and Wiest, Roland and Kirschke, Jan and Wiestler, Benedikt and Colen, Rivka and Kotrotsou, Aikaterini and Lamontagne, Pamela and Marcus, Daniel and Milchenko, Mikhail and Nazeri, Arash and Weber, Marc-Andre and Mahajan, Abhishek and Baid, Ujjwal and Gerstner, Elizabeth and Kwon, Dongjin and Acharya, Gagan and Agarwal, Manu and Alam, Mahbubul and Albiol, Alberto and Albiol, Antonio and Albiol, Francisco J. and Alex, Varghese and Allinson, Nigel and Amorim, Pedro H. A. and Amrutkar, Abhijit and Anand, Ganesh and Andermatt, Simon and Arbel, Tal and Arbelaez, Pablo and Avery, Aaron and Azmat, Muneeza and B., Pranjal and Bai, W and Banerjee, Subhashis and Barth, Bill and Batchelder, Thomas and Batmanghelich, Kayhan and Battistella, Enzo and Beers, Andrew and Belyaev, Mikhail and Bendszus, Martin and Benson, Eze and Bernal, Jose and Bharath, Halandur Nagaraja and Biros, George and Bisdas, Sotirios and Brown, James and Cabezas, Mariano and Cao, Shilei and Cardoso, Jorge M. and Carver, Eric N and Casamitjana, Adrià and Castillo, Laura Silvana and Catà, Marcel and Cattin, Philippe and Cerigues, Albert and Chagas, Vinicius S. and Chandra, Siddhartha and Chang, Yi-Ju and Chang, Shiyu and Chang, Ken and Chazalon, Joseph and Chen, Shengcong and Chen, Wei and Chen, Jefferson W and Chen, Zhaolin and Cheng, Kun and Choudhury, Ahana Roy and Chylla, Roger and Clérigues, Albert and Colleman, Steven and Colmeiro, Ramiro German Rodriguez and Combalia, Marc and Costa, Anthony and Cui, Xiaomeng and Dai, Zhenzhen and Dai, Lutao and Daza, Laura Alexandra and Deutsch, Eric and Ding, Changxing and Dong, Chao and Dong, Shidu and Dudzik, Wojciech and Eaton-Rosen, Zach and Egan, Gary and Escudero, Guilherme and Estienne, Théo and Everson, Richard and Fabrizio, Jonathan and Fan, Yong and Fang, Longwei and Feng, Xue and Ferrante, Enzo and Fidon, Lucas and Fischer, Martin and French, Andrew P. and Fridman, Naomi and Fu, Huan and Fuentes, David and Gao, Yaozong and Gates, Evan and Gering, David and Gholami, Amir and Gierke, Willi and Glocker, Ben and Gong, Mingming and González-Villá, Sandra and Grosges, T. and Guan, Yuanfang and Guo, Sheng and Gupta, Sudeep and Han, Woo-Sup and Han, Il Song and Harmuth, Konstantin and He, Huiguang and Hernández-Sabaté, Aura and Herrmann, Evelyn and Himthani, Naveen and Hsu, Winston and Hsu, Cheyu and Hu, Xiaojun and Hu, Xiaobin and Hu, Yan and Hu, Yifan and Hua, Rui and Huang, Teng-Yi and Huang, Weilin and Van Huffel, Sabine and Huo, Quan and HV, Vivek and Iftekharuddin, Khan M. and Isensee, Fabian and Islam, Mobarakol and Jackson, Aaron S. and Jambawalikar, Sachin R. and Jesson, Andrew and Jian, Weijian and Jin, Peter and Jose, V Jeya Maria and Jungo, Alain and Kainz, B and Kamnitsas, Konstantinos and Kao, Po-Yu and Karnawat, Ayush and Kellermeier, Thomas and Kermi, Adel and Keutzer, Kurt and Khadir, Mohamed Tarek and Khened, Mahendra and Kickingereder, Philipp and Kim, Geena and King, Nik and Knapp, Haley and Knecht, Urspeter and Kohli, Lisa and Kong, Deren and Kong, Xiangmao and Koppers, Simon and Kori, Avinash and Krishnamurthi, Ganapathy and Krivov, Egor and Kumar, Piyush and Kushibar, Kaisar and Lachinov, Dmitrii and Lambrou, Tryphon and Lee, Joon and Lee, Chengen and Lee, Yuehchou and Lee, M and Lefkovits, Szidonia and Lefkovits, Laszlo and Levitt, James and Li, Tengfei and Li, Hongwei and Li, Wenqi and Li, Hongyang and Li, Xiaochuan and Li, Yuexiang and Li, Heng and Li, Zhenye and Li, Xiaoyu and Li, Zeju and Li, XiaoGang and Li, Wenqi and Lin, Zheng-Shen and Lin, Fengming and Lio, Pietro and Liu, Chang and Liu, Boqiang and Liu, Xiang and Liu, Mingyuan and Liu, Ju and Liu, Luyan and Llado, Xavier and Lopez, Marc Moreno and Lorenzo, Pablo Ribalta and Lu, Zhentai and Luo, Lin and Luo, Zhigang and Ma, Jun and Ma, Kai and Mackie, Thomas and Madabushi, Anant and Mahmoudi, Issam and Maier-Hein, Klaus H. and Maji, Pradipta and Mammen, CP and Mang, Andreas and Manjunath, B. S. and Marcinkiewicz, Michal and McDonagh, S and McKenna, Stephen and McKinley, Richard and Mehl, Miriam and Mehta, Sachin and Mehta, Raghav and Meier, Raphael and Meinel, Christoph and Merhof, Dorit and Meyer, Craig and Miller, Robert and Mitra, Sushmita and Moiyadi, Aliasgar and Molina-Garcia, David and Monteiro, Miguel A. B. and Mrukwa, Grzegorz and Myronenko, Andriy and Nalepa, Jakub and Ngo, Thuyen and Nie, Dong and Ning, Holly and Niu, Chen and Nuechterlein, Nicholas K and Oermann, Eric and Oliveira, Arlindo and Oliveira, Diego D. C. and Oliver, Arnau and Osman, Alexander F. I. and Ou, Yu-Nian and Ourselin, Sebastien and Paragios, Nikos and Park, Moo Sung and Paschke, Brad and Pauloski, J. Gregory and Pawar, Kamlesh and Pawlowski, Nick and Pei, Linmin and Peng, Suting and Pereira, Silvio M. and Perez-Beteta, Julian and Perez-Garcia, Victor M. and Pezold, Simon and Pham, Bao and Phophalia, Ashish and Piella, Gemma and Pillai, G. N. and Piraud, Marie and Pisov, Maxim and Popli, Anmol and Pound, Michael P. and Pourreza, Reza and Prasanna, Prateek and Prkovska, Vesna and Pridmore, Tony P. and Puch, Santi and Puybareau, Élodie and Qian, Buyue and Qiao, Xu and Rajchl, Martin and Rane, Swapnil and Rebsamen, Michael and Ren, Hongliang and Ren, Xuhua and Revanuru, Karthik and Rezaei, Mina and Rippel, Oliver and Rivera, Luis Carlos and Robert, Charlotte and Rosen, Bruce and Rueckert, Daniel and Safwan, Mohammed and Salem, Mostafa and Salvi, Joaquim and Sanchez, Irina and Sánchez, Irina and Santos, Heitor M. and Sartor, Emmett and Schellingerhout, Dawid and Scheufele, Klaudius and Scott, Matthew R. and Scussel, Artur A. and Sedlar, Sara and Serrano-Rubio, Juan Pablo and Shah, N. Jon and Shah, Nameetha and Shaikh, Mazhar and Shankar, B. Uma and Shboul, Zeina and Shen, Haipeng and Shen, Dinggang and Shen, Linlin and Shen, Haocheng and Shenoy, Varun and Shi, Feng and Shin, Hyung Eun and Shu, Hai and Sima, Diana and Sinclair, M and Smedby, Orjan and Snyder, James M. and Soltaninejad, Mohammadreza and Song, Guidong and Soni, Mehul and Stawiaski, Jean and Subramanian, Shashank and Sun, Li and Sun, Roger and Sun, Jiawei and Sun, Kay and Sun, Yu and Sun, Guoxia and Sun, Shuang and Suter, Yannick R and Szilagyi, Laszlo and Talbar, Sanjay and Tao, Dacheng and Tao, Dacheng and Teng, Zhongzhao and Thakur, Siddhesh and Thakur, Meenakshi H and Tharakan, Sameer and Tiwari, Pallavi and Tochon, Guillaume and Tran, Tuan and Tsai, Yuhsiang M. and Tseng, Kuan-Lun and Tuan, Tran Anh and Turlapov, Vadim and Tustison, Nicholas and Vakalopoulou, Maria and Valverde, Sergi and Vanguri, Rami and Vasiliev, Evgeny and Ventura, Jonathan and Vera, Luis and Vercauteren, Tom and Verrastro, C. A. and Vidyaratne, Lasitha and Vilaplana, Veronica and Vivekanandan, Ajeet and Wang, Guotai and Wang, Qian and Wang, Chiatse J. and Wang, Weichung and Wang, Duo and Wang, Ruixuan and Wang, Yuanyuan and Wang, Chunliang and Wang, Guotai and Wen, Ning and Wen, Xin and Weninger, Leon and Wick, Wolfgang and Wu, Shaocheng and Wu, Qiang and Wu, Yihong and Xia, Yong and Xu, Yanwu and Xu, Xiaowen and Xu, Peiyuan and Yang, Tsai-Ling and Yang, Xiaoping and Yang, Hao-Yu and Yang, Junlin and Yang, Haojin and Yang, Guang and Yao, Hongdou and Ye, Xujiong and Yin, Changchang and Young-Moxon, Brett and Yu, Jinhua and Yue, Xiangyu and Zhang, Songtao and Zhang, Angela and Zhang, Kun and Zhang, Xuejie and Zhang, Lichi and Zhang, Xiaoyue and Zhang, Yazhuo and Zhang, Lei and Zhang, Jianguo and Zhang, Xiang and Zhang, Tianhao and Zhao, Sicheng and Zhao, Yu and Zhao, Xiaomei and Zhao, Liang and Zheng, Yefeng and Zhong, Liming and Zhou, Chenhong and Zhou, Xiaobing and Zhou, Fan and Zhu, Hongtu and Zhu, Jin and Zhuge, Ying and Zong, Weiwei and Kalpathy-Cramer, Jayashree and Farahani, Keyvan and Davatzikos, Christos and van Leemput, Koen and Menze, Bjoern},
  year = {2019},
  eprint = {1811.02629v3},
  archiveprefix = {arXiv},
  primaryclass = {cs.CV},
  url = {https://arxiv.org/abs/1811.02629v3},
  doi = {10.48550/arXiv.1811.02629},
  howpublished = {arXiv preprint arXiv:1811.02629v3},
}

@inproceedings{wu2026codebrain,
  author = {Wu, Yicheng and Song, Tao and Wu, Zhonghua and Ye, Jin and Ge, Zongyuan and Bai, Wenjia and Chen, Zhaolin and Cai, Jianfei},
  title = {{Virtual Full-stack Scanning of Brain MRI via Imputing Any Quantised Code}},
  booktitle = {Proceedings of the IEEE/CVF Conference on Computer Vision and Pattern Recognition (CVPR)},
  year = {2026},
  pages = {21026-21035},
  url = {https://openaccess.thecvf.com/content/CVPR2026/html/Wu_Virtual_Full-stack_Scanning_of_Brain_MRI_via_Imputing_Any_Quantised_CVPR_2026_paper.html},
}

@article{diffm4ri,
  author = {Ye, Wen and Guo, Zhetao and Ren, Yuxiang and Tian, Yi and Shen, Yushi and Chen, Zan and He, Junjun and Ke, Jing and Shen, Yiqing},
  journal = {IEEE Journal of Biomedical and Health Informatics},
  title = {{DiffM$^{4}$RI: A Latent Diffusion Model With Modality Inpainting for Synthesizing Missing Modalities in MRI Analysis}},
  year = {2026},
  volume = {30},
  number = {2},
  pages = {1006--1018},
  doi = {10.1109/jbhi.2025.3580510},
  url = {https://doi.org/10.1109/jbhi.2025.3580510},
}

@article{m2dn,
  author = {Meng, Xiangxi and Sun, Kaicong and Xu, Jun and He, Xuming and Shen, Dinggang},
  journal = {IEEE Transactions on Medical Imaging},
  title = {{Multi-Modal Modality-Masked Diffusion Network for Brain MRI Synthesis With Random Modality Missing}},
  year = {2024},
  volume = {43},
  number = {7},
  pages = {2587--2598},
  doi = {10.1109/tmi.2024.3368664},
  url = {https://doi.org/10.1109/tmi.2024.3368664},
}

@article{mmgan,
  author = {Sharma, Anmol and Hamarneh, Ghassan},
  journal = {IEEE Transactions on Medical Imaging},
  title = {{Missing MRI Pulse Sequence Synthesis Using Multi-Modal Generative Adversarial Network}},
  year = {2020},
  volume = {39},
  number = {4},
  pages = {1170--1183},
  doi = {10.1109/tmi.2019.2945521},
  url = {https://doi.org/10.1109/tmi.2019.2945521},
}

@article{mmsyn,
  author = {Chartsias, Agisilaos and Joyce, Thomas and Giuffrida, Mario Valerio and Tsaftaris, Sotirios A.},
  journal = {IEEE Transactions on Medical Imaging},
  title = {{Multimodal MR Synthesis via Modality-Invariant Latent Representation}},
  year = {2018},
  volume = {37},
  number = {3},
  pages = {803--814},
  doi = {10.1109/tmi.2017.2764326},
  url = {https://doi.org/10.1109/tmi.2017.2764326},
}

@inproceedings{ummi,
  author = {Qi, Liangce and Liu, Yusi and Li, Yuqin and Shi, Weili and Feng, Guanyuan and Jiang, Zhengang},
  title = {{A Unified Missing Modality Imputation Model with Inter-modality Contrastive and Consistent Learning}},
  booktitle = {Medical Image Computing and Computer Assisted Intervention -- MICCAI 2025},
  year = {2025},
  publisher = {Springer Nature Switzerland},
  volume = {15967},
  page = {44 -- 53},
  doi = {10.1007/978-3-032-04984-1_5},
  url = {https://doi.org/10.1007/978-3-032-04984-1_5},
  pages = {44--53},
  series = {Lecture Notes in Computer Science},
}

@article{Birrer_2018,
  title = {{lenstronomy: Multi-purpose gravitational lens modelling software package}},
  volume = {22},
  issn = {2212-6864},
  url = {https://doi.org/10.1016/j.dark.2018.11.002},
  doi = {10.1016/j.dark.2018.11.002},
  journal = {Physics of the Dark Universe},
  publisher = {Elsevier BV},
  author = {Birrer, Simon and Amara, Adam},
  year = {2018},
  pages = {189--201},
}

@inproceedings{repaint,
  title = {{RePaint: Inpainting using Denoising Diffusion Probabilistic Models}},
  author = {Lugmayr, Andreas and Danelljan, Martin and Romero, Andres and Yu, Fisher and Timofte, Radu and Van Gool, Luc},
  year = {2022},
  url = {https://doi.org/10.1109/cvpr52688.2022.01117},
  doi = {10.1109/cvpr52688.2022.01117},
  pages = {11461--11471},
  booktitle = {2022 IEEE/CVF Conference on Computer Vision and Pattern Recognition (CVPR)},
  publisher = {IEEE},
}

@inproceedings{lpips,
  title = {{The Unreasonable Effectiveness of Deep Features as a Perceptual Metric}},
  author = {Zhang, Richard and Isola, Phillip and Efros, Alexei A. and Shechtman, Eli and Wang, Oliver},
  booktitle = {2018 IEEE/CVF Conference on Computer Vision and Pattern Recognition},
  pages = {586--595},
  year = {2018},
  doi = {10.1109/cvpr.2018.00068},
  url = {https://doi.org/10.1109/cvpr.2018.00068},
}

@article{schawinski2017generative,
  title = {{Generative adversarial networks recover features in astrophysical images of galaxies beyond the deconvolution limit}},
  author = {Schawinski, Kevin and Zhang, Ce and Zhang, Hantian and Fowler, Lucas and Santhanam, Gokula Krishnan},
  journal = {Monthly Notices of the Royal Astronomical Society: Letters},
  volume = {467},
  number = {1},
  pages = {L110--L114},
  year = {2017},
  doi = {10.1093/mnrasl/slx008},
  url = {https://doi.org/10.1093/mnrasl/slx008},
}

@article{miao2024astrosr,
  title = {{AstroSR: A Data Set of Galaxy Images for Astronomical Superresolution Research}},
  author = {Miao, Jiawei and Tu, Liangping and Jiang, Bin and Li, Xiangru and Qiu, Bo},
  journal = {The Astrophysical Journal Supplement Series},
  volume = {274},
  number = {1},
  pages = {7},
  year = {2024},
  doi = {10.3847/1538-4365/ad61e4},
  url = {https://doi.org/10.3847/1538-4365/ad61e4},
}

@article{reddy2024difflense,
  doi = {10.1088/2632-2153/ad76f8},
  url = {https://doi.org/10.1088/2632-2153/ad76f8},
  year = {2024},
  publisher = {IOP Publishing},
  volume = {5},
  number = {3},
  pages = {035076},
  author = {Reddy, Pranath and Toomey, Michael W and Parul, Hanna and Gleyzer, Sergei},
  title = {{DiffLense: a conditional diffusion model for super-resolution of gravitational lensing data}},
  journal = {Machine Learning: Science and Technology},
}

@inproceedings{wu2026star,
  title = {{STAR: A Benchmark for Astronomical Star Fields Super-Resolution}},
  author = {Wu, Guocheng and Zhuang, Guohang and Huang, Jinyang and Zhang, Xiang and Ouyang, Wanli and Lu, Yan},
  booktitle = {Advances in Neural Information Processing Systems},
  year = {2025},
  url = {https://doi.org/10.52202/085713-4765},
  doi = {10.52202/085713-4765},
  pages = {157998--158033},
  volume = {38},
  publisher = {Curran Associates, Inc.},
}

@inproceedings{bouman2016computational,
  title = {{Computational Imaging for VLBI Image Reconstruction}},
  author = {Bouman, Katherine L. and Johnson, Michael D. and Zoran, Daniel and Fish, Vincent L. and Doeleman, Sheperd S. and Freeman, William T.},
  booktitle = {2016 IEEE Conference on Computer Vision and Pattern Recognition (CVPR)},
  year = {2016},
  doi = {10.1109/cvpr.2016.105},
  url = {https://doi.org/10.1109/cvpr.2016.105},
  pages = {913--922},
}

@article{bouman2017reconstructing,
  author = {Bouman, Katherine L. and Johnson, Michael D. and Dalca, Adrian V. and Chael, Andrew A. and Roelofs, Freek and Doeleman, Sheperd S. and Freeman, William T.},
  title = {{Reconstructing Video of Time-Varying Sources From Radio Interferometric Measurements}},
  year = {2018},
  cdate = {1514764800000},
  journal = {IEEE Transactions on Computational Imaging},
  volume = {4},
  number = {4},
  pages = {512--527},
  url = {https://doi.org/10.1109/tci.2018.2838452},
  doi = {10.1109/tci.2018.2838452},
}

@article{conselice2014evolution,
  title = {{The Evolution of Galaxy Structure Over Cosmic Time}},
  volume = {52},
  issn = {1545-4282},
  url = {https://doi.org/10.1146/annurev-astro-081913-040037},
  doi = {10.1146/annurev-astro-081913-040037},
  number = {1},
  journal = {Annual Review of Astronomy and Astrophysics},
  publisher = {Annual Reviews},
  author = {Conselice, Christopher J.},
  year = {2014},
  pages = {291--337},
}

@article{appleton1996collisional,
  author = {Appleton, P. N. and Struck-Marcell, Curtis},
  title = {{Collisional Ring Galaxies}},
  journal = {Fundamentals of Cosmic Physics},
  year = {1996},
  volume = {16},
  pages = {111--220},
  adsurl = {https://ui.adsabs.harvard.edu/abs/1996FCPh...16..111A},
  url = {https://ned.ipac.caltech.edu/level5/Sept01/Appleton/Appleton_contents.html},
}

@article{walmsley2022galaxy,
  title = {{Galaxy Zoo DECaLS: Detailed visual morphology measurements from volunteers and deep learning for 314 000 galaxies}},
  volume = {509},
  issn = {1365-2966},
  url = {https://doi.org/10.1093/mnras/stab2093},
  doi = {10.1093/mnras/stab2093},
  number = {3},
  journal = {Monthly Notices of the Royal Astronomical Society},
  publisher = {Oxford University Press (OUP)},
  author = {Walmsley, Mike and Lintott, Chris and Géron, Tobias and Kruk, Sandor and Krawczyk, Coleman and Willett, Kyle W and Bamford, Steven and Kelvin, Lee S and Fortson, Lucy and Gal, Yarin and Keel, William and Masters, Karen L and Mehta, Vihang and Simmons, Brooke D and Smethurst, Rebecca and Smith, Lewis and Baeten, Elisabeth M and Macmillan, Christine},
  year = {2022},
  pages = {3966--3988},
}


\appendix

\section{Galaxy--Galaxy Strong Lensing}
\label{app:strong-lensing}

Strong gravitational lensing occurs when a massive foreground galaxy, referred to as the deflector or lens, lies close to the line of sight to a more distant background source such that bending of light due to the foreground mass results in multiple images of the same background source. 
The gravitational potential of the deflector bends the light from the background source and changes the apparent position, shape, and brightness of the source image. 
In the thin-lens approximation, the relation between the angular position of the source $\boldsymbol{\beta}$ and the observed image position $\boldsymbol{\theta}$ is given by the lens equation
\begin{equation}
    \boldsymbol{\beta}
    =
    \boldsymbol{\theta}
    -
    \boldsymbol{\alpha}(\boldsymbol{\theta}),
\end{equation}
where $\boldsymbol{\beta}$ and $\boldsymbol{\theta}$ are two-dimensional angular coordinates on the source and image planes, and $\boldsymbol{\alpha}$ is the reduced deflection angle.
When the projected source position lies sufficiently close to the optical-axis the lens equation can have multiple solutions, corresponding to multiple observed images of the same background source. For a more detailed introduction into strong gravitational lensing, we refer to \cite{Saha:2024}.

For an axisymmetric lens, the characteristic angular scale of strong lensing is the Einstein radius. 
For a singular isothermal sphere, it can be written as
\begin{equation}
    \theta_{\rm E}
    =
    4\pi
    \left(\frac{\sigma_v}{c}\right)^2
    \frac{D_{\rm ls}}{D_{\rm s}},
\end{equation}
where $\sigma_v$ is the velocity dispersion of the lens, $c$ is the speed of light, $D_{\rm s}$ is the angular-diameter distance to the source, and $D_{\rm ls}$ is the angular-diameter distance between the lens and the source. For a typical galaxy as a deflector, the Einstein radius is of order $0.1-2$ arc seconds, right in the regime of (partial) blending for ground-based observations. 
When the source is nearly aligned with the deflector, its light can be highly magnified and stretched tangentially around the lens center, producing multiple images, arcs, or nearly complete Einstein rings~\cite{treu2010strong, kochanek2006strong}. 
For extended background galaxies, the observed lensing morphology is therefore not a simple compact source but a non-local structure distributed around the foreground deflector.

These arc-like structures are scientifically important because their positions, curvatures, thicknesses, and surface-brightness distributions encode information about the projected mass distribution of the deflector, dark-matter substructure, and the morphology of the high-redshift background source (e.g., \cite{treu2010strong, Shajib:2024, Vegetti:2024} for a set of reviews).

Galaxy--galaxy strong lenses are morphologically distinct from ordinary galaxies. 
Typical isolated galaxies often have smooth and centrally concentrated light profiles, whereas strong-lensing systems contain a superposition of the foreground deflector and one or more distorted images of a background source. 
The lensed arcs are frequently faint, spatially extended, low-surface-brightness features that may be partially blended with the deflector light. 
Their visibility depends strongly on survey-specific point-spread functions, pixel scales, noise levels, exposure depths, and bandpasses. 
Consequently, the same physical lens system can appear substantially different across LSST, Euclid, and Roman observations.

Strong lenses are intrinsically rare because they require a special geometric configuration: the background source must fall within a small lensing cross-section around the foreground deflector. 
In wide-field surveys, only a tiny fraction of galaxies exhibit recognizable strong-lensing configurations. 
Current and upcoming surveys are expected to detect on the order of $10^4$--$10^5$ strong lenses among $\sim10^{9}$--$10^{10}$ galaxies, corresponding to an occurrence rate of roughly $10^{-5}$--$10^{-4}$ per galaxy~\cite{collett2015population, oguri2010statistics}. 
Because of this rarity, strong-lens discovery has been widely studied using machine learning methods for identifying lens candidates~\cite{lanusse2018cmu, jacobs2019finding}.

For cross-survey imputation, galaxy--galaxy strong lenses provide a stringent rare-event test case. 
A model trained primarily on common galaxy populations may learn a strong prior toward smooth, centrally concentrated profiles. 
When evaluated on strong lenses, such a model may suppress low-surface-brightness arcs, break continuous arcs into compact clumps, hallucinate regular galaxy structure, or fail to preserve the angular extent of the lensed source. 
This failure mode is especially important for scientific use: an imputed image can obtain reasonable pixel-level reconstruction metrics while still destroying the lensing morphology needed for downstream analysis.

In our benchmark, we therefore use galaxy--galaxy strong lenses as an out-of-distribution rare-event evaluation set. 
Models are trained on common galaxy populations and evaluated zero-shot on strong-lensing systems. 
This setting mimics early-stage survey deployment, where rare objects may be observed in existing surveys but remain sparsely covered by a new target survey. 
It allows us to evaluate whether cross-survey imputation methods can preserve scientifically meaningful rare morphology, rather than merely reconstructing the average appearance of ordinary galaxies.

\section{Data Generator Implementation Details}
\label{app:data-gen}
We construct a physically forward-modelled multi-survey dataset for cross-survey generation using \texttt{slsim}~\cite{khadka2026slsim}, an open-source astronomical object and imaging simulation package, together with \texttt{lenstronomy}~\cite{Birrer_2018}, a strong gravitational lensing simulation package. The pipeline separates population generation from image simulation: galaxy populations are drawn with \texttt{SkyPy}~\cite{Amara2021}, a software package simulating catalog-level galaxy populations calibrated to astronomical surveys catalogs, while lensing calculations, point spread function (PSF) convolution, pixelization and noise rendering are handled downstream. Each galaxy has a description of its spectral energy density, such that flux can be computed for any of the band pass filters of the different instruments self consistently. Throughout this work, we adopt a concordance cosmological model with a flat geometry, a cosmological constant $\Lambda$ and cold dark matter ($\Lambda$CDM). Within this model we set the Hubble constant $H_0 = 70\ \mathrm{km\ s^{-1}\ Mpc^{-1}}$ and the current-time matter density $\Omega_m = 0.3$.


\subsection{Survey Setup}

Galaxy populations are built from a shared \texttt{SkyPy} configuration including spectral energy densities. The filter troughputs for the different instruments and filter choices are managed through \texttt{speclite}~\cite{speclite_zenodo_2024}. We render Euclid in VIS, LSST in $g,r,i,z,y$, and Roman in F106($Y$), F129($J$), and F158($H$). These filters are not all confined to the visible regime: Euclid VIS and the LSST optical filters mainly cover the visible/optical range with some extension toward the near-infrared, whereas the Roman F106, F129, and F158 filters are near-infrared bands. The Roman band choice follows the nominal High-Latitude Wide-Area Survey (HLWAS) imaging configuration, in which these bands define the core near-infrared multi-band setup.

To preserve survey-native angular sampling and maintain an approximately consistent field of view across surveys, we choose survey-specific cutout sizes based on the native pixel scale of each instrument instead of enforcing a common pixel grid. In the current dataset release, we use cutout sizes of $64\times64$ for Euclid, $32\times32$ for LSST, and $59\times59$ for Roman, yielding approximately matched angular coverage across the three surveys while retaining their native sampling characteristics. 

\subsection{Common Galaxy Population}

The common set is built from the same parent \texttt{SkyPy}-based catalog family used for the lensing pipeline. We retain two physically interpretable classes, namely red early-type galaxies and blue late-type galaxies, and apply simple astrophysical cuts to keep the common and lens samples consistent at the level of cosmology and photometric assumptions. In particular, we filter deflector-like galaxies with astronomical magnitude limits in the $i$-band $i<24$ and in the cosmological redshift range $0.01<z<2$, and source-like galaxies with $i<26$ and $0.01<z<5$.

To increase morphological realism, non-lens scenes may include projected neighboring galaxies introduced by Poisson sampling the sky density of galaxies from our catalog with random positions on the sky. This design introduces realistic crowding, blending, and lens-like false positives while preserving the non-lensing label. In the dataset, the non-lens manifests contain 115{,}000 examples across eight runs, resulting in 71.49 GB of data on disk.

\subsection{Strong-lens Population}

The strong-lens set is constructed from the same cosmology and parent galaxy catalogs, but with explicit separation between foreground deflectors and background sources before constructing galaxy--galaxy lenses with \texttt{SLSim}'s \texttt{LensPop} class. Deflectors are selected with $0.1<z_d<2$ and $i<24$, while source galaxies are drawn from the blue galaxy population with $0.1<z_s<5$ and $i<26$. The light distribution of the sources are modeled as single-S\'ersic emitters~\cite{Sersic:1963}.

We retain only systems satisfying detectability-oriented cuts: image separations between $0.1$ and $8.0$ arcsec, and a second-brightest lensed image brighter than $i=25$. In the dataset, each run begins from a cached population draw and writes accepted systems into sharded Zarr files. 
The two completed runs contain 14{,}825 accepted lens systems in total, resulting in 13.36 GB of data on disk.



\section{Experimental Details}
\label{app:training-details}
\subsection{Native-Grid Correspondence and Cross-Resolution Imputation}
\label{app:native-grid}
All survey observations are forward-rendered around a common angular center. They are therefore co-centered but not pixelwise aligned: the center-and-scale mapping between survey grids is fixed and shared across the dataset. The task remains imputation of an unobserved survey modality, even when source and target pixel scales differ.

Native MRI acquisitions can also differ in spatial resolution, although BraTS-style benchmarks resample them to a common isotropic grid~\cite{menze2015brats,bakas2018identifying}. COSMIO deliberately retains survey-native pixel scales. Padding leaves the observed pixels and their native noise structure unchanged, whereas common-grid resampling interpolates neighboring pixels and alters noise correlations. This distinction is particularly relevant for noisy survey imaging. Table~\ref{tab:preprocessing-ablation} compares padding with common-grid rescaling for regular-galaxy Roman imputation. Padding gives lower RMSE and CRPS and higher PSNR in this native-grid evaluation.
\begin{table}[t]
\centering\small
\caption{Padding versus common-grid rescaling for regular-galaxy Roman imputation, evaluated on the native target grid. Lower is better for RMSE and CRPS; higher is better for PSNR.}
\label{tab:preprocessing-ablation}
\begin{tabular}{lrrr}
\toprule
Preprocessing & RMSE$\downarrow$ & PSNR $\uparrow$ & CRPS$\downarrow$ \\
\midrule
Padding & \textbf{0.184} & \textbf{39.190} & \textbf{0.069} \\
Rescaling & 0.277 & 36.105 & 0.074 \\
\bottomrule
\end{tabular}
\end{table}

\subsection{Brain-MRI Imputation Baselines}
\label{app:mri-architecture-details}

The methods adapted from brain MRI imputation followed the model families
specified in their original papers, with only the input/output modality
interfaces changed to match the surveys.  In all cases, the medical-image
modalities were replaced by LSST, Euclid, and Roman tensors.

\paragraph{CodeBrain.}
CodeBrain was trained as a two-stage model over the packed 9-channel
LSST--Euclid--Roman tensor.  The model used quantization level $L=5$, code
dimension 9, common feature dimension 128, and base convolutional width 64.
Stage I trained the source encoder, posterior code encoder, decoder, and
PatchGAN discriminator for reconstruction with quantized anchor codes.  Stage
II froze the Stage-I reconstruction modules and trained the prior encoder to
predict full-modality quantized codes from incomplete inputs.

\paragraph{DiffM4RI.}
DiffM4RI was implemented as a 2D latent-diffusion model.  A multimodal VQ-VAE
first encoded the packed survey tensor into a latent representation, and a
DiT-style denoiser was then trained unconditionally in the latent space.  The
VQ-VAE used 32 latent channels, hidden width 128, two residual blocks, codebook
size 1024, commitment weight 0.25, and codebook loss weight 1.0.

The latent diffusion denoiser used hidden size 512, depth 12, 8 attention
heads, patch size 2, MLP ratio 4.0, and learned variance prediction.  The
diffusion process used 1000 timesteps with the shared linear beta schedule.

\paragraph{UMMI.}
UMMI used the dual-branch encoder architecture from the unified
missing-modality imputation model with inter-modality contrastive and
consistency learning.  The model used base dimension 96, stage depths
$(2,2,2,2)$, attention heads $(4,6,8,12)$, window size 4, patch size 2, MLP
ratio 4.0, and projection dimension 512.  The loss combined reconstruction,
inter-modality contrastive, and consistency terms.  The contrastive loss weight
was 0.3, the consistency loss weight was 0.5, the contrastive temperature was
0.8, and the consistency temperature was 0.9. The spatial mask ratio was 0.75.

\paragraph{MMSyn.}
MMSyn used per-modality encoders, a modality-invariant latent representation,
pixel-wise maximum latent fusion, and per-output decoder heads.  The
implementation used 16 latent channels and LeakyReLU slope 0.1.  Each batch
optimized the three MMSyn objective components: single-modality decode
consistency, latent similarity regularization, and fused-latent multimodal
reconstruction.  The three loss weights were all set to 1.0.

\paragraph{M2DN.}
M2DN was trained as a modality-masked diffusion model for Roman synthesis from
incomplete LSST/Euclid/Roman inputs.  The model used base
channels 48, fusion channels 128, bottleneck channels 192, and the shared
1000-step linear diffusion schedule.  Masked inputs were filled with 0.
\paragraph{MMGAN.}
MMGAN was trained as a conditional multimodal GAN for missing-modality
synthesis. The generator took the observed survey stack with missing channels
zeroed, modality-presence indicators, latent noise, and a learned cluster
embedding. Random observed-mask training sampled non-trivial modality subsets
with at least one observed and one missing modality. The model used 8 mixture
clusters, noise dimension 64, minimum mixture standard deviation 0.1, generator
base width 128, discriminator and encoder base width 64, and cluster embedding
dimension 16. The objective combined the cRaSGAN adversarial loss, cluster
cross-entropy with weight 1.0, and missing-modality reconstruction with weight
100.0.

Overview of methods are shown in Table~\ref{tab:baseline_methods}.
\begin{table}[t]

\caption{Overview of the brain-MRI and physically motivated imputation baselines.}
\label{tab:baseline_methods}
\centering
\small
\begin{tabular}{lccc}
\toprule
Method & Year & Core Idea & Source \\
\midrule
CodeBrain~\cite{wu2026codebrain} & 2026 & Quantized-code imputation & CVPR \\
DiffM4RI~\cite{diffm4ri}  & 2026 & Latent DiT inpainting & IEEE JBHI \\
MMGAN~\cite{mmgan}     & 2020 & GAN + GMM & IEEE TMI \\
UMMI~\cite{ummi}      & 2025 & Mask-pretrained transformer & MICCAI \\
MMSyn~\cite{mmsyn}     & 2018 & Modality-invariant latent & IEEE TMI \\
M2DN~\cite{m2dn}      & 2024 & Latent Conditional Diffusion & IEEE TMI \\
\midrule
PSF-Linear & & PSF matching + ridge color map & \\
S\'ersic-FM & & Shared S\'ersic fit + forward rendering & \\
\bottomrule
\end{tabular}
\end{table}

\subsection{Diffusion Baselines}
\label{app:diffusion-baselines}

For a fair comparison, the three direct diffusion baselines used the same
simple CNN denoising backbone:

An initial 3×3 convolution; three resolution stages with channel multipliers (1, 2, 4) from a base width of 64; two residual blocks per stage.

Each residual block includes:

GroupNorm and SiLU activation; two 3×3 convolutions; FiLM-style timestep conditioning via a learned projection of the time embedding.

The bottleneck contains two additional residual blocks. Downsampling is implemented via stride-2 convolutions, and upsampling via nearest-neighbor interpolation followed by convolution. The decoder uses standard U-Net skip connections.  All three used epsilon prediction, clipped
$x_0$ estimates, and the shared 1000-step linear beta schedule.

RePaint trained a single DDPM over the concatenated 9-channel
LSST--Euclid--Roman tensor. Palette trained target-specific DDPMs conditioned on one or more
source modalities.

Asyn-Diffusion trained a joint asynchronous diffusion model over the
concatenated 9-channel tensor, sharing one model across all modalities and
imputation patterns. The denoiser used the shared three-level U-Net with
three modality-specific timestep inputs. These inputs encode modality
availability directly, without a separate modality embedding or specialized
MRI modules such as channel attention. This minimal design isolates
modality-level conditioning within the shared diffusion backbone.

The modality-specific timestep embeddings are concatenated and provided to the denoiser. During sampling, observed modalities are assigned timestep zero, corresponding to clean inputs, while missing modalities are progressively denoised from higher noise levels to zero.

\subsection{Physically Motivated Baselines}
\label{app:physical-baselines}
We evaluate two deterministic physical baselines for LSST+Euclid $\rightarrow$ Roman imputation.

\paragraph{PSF-Linear.}
We convert the five LSST bands and Euclid VIS to physical flux units and flux-conservingly resample them onto the Roman grid. Each input band is convolved with a band-specific Gaussian kernel. The Roman training targets are homogenized to the same PSF. We then fit a global ridge-regression color transform, shared across all objects and pixels, from the six input-band fluxes to the three Roman-band fluxes. The entire model contains only a $6\times3$ coefficient matrix and three intercepts.

\paragraph{S\'ersic-FM.}
We jointly fit a single elliptical S\'ersic morphology to all six input bands while accounting for their survey-specific PSFs, native pixel grids, and noise variances. The shared morphology contains six nonlinear parameters: the two-dimensional center, effective radius, S\'ersic index, axis ratio, and position angle. The six band amplitudes are solved by weighted linear least squares conditional on the morphology. We then re-render the fitted profile through each Roman-band PSF on the native Roman grid. Roman-band total fluxes are predicted from the fitted input-band fluxes using a global ridge map learned exclusively from the training split. Unlike PSF-Linear, S\'ersic-FM produces native-resolution Roman predictions without PSF homogenization.

Since both methods are deterministic, we report their RMSE and PSNR without CRPS, together with downstream lens classification AUROC and AUPRC. The supplied reconstruction results are reported in Tables~\ref{tab:main_results_full} and~\ref{tab:lens_flux_results_full}, and the classification results in Table~\ref{tab:downstream-classification-auroc-auprc}.

\subsection{Downstream Lens Classifier}
\label{app:training-details:classifier}

The downstream classifier was trained as a Roman-only binary lens classifier
for measuring how imputation quality affects the lens-versus-non-lens task.
Each input was a 3-channel Roman image resized to $224 \times 224$.  The model
used the torchvision VisionTransformer implementation with patch size 16, 6
transformer layers, 8 attention heads, hidden dimension 256, MLP dimension 512,
zero dropout, zero attention dropout, and 2 output classes.

\subsection{Downstream Evaluation Protocol and Lens Prevalence}
\label{app:lens-prior}
For each method, the Roman lens classifier is trained on imputed lens and non-lens images and evaluated directly on held-out ground-truth Roman images, with disjoint object identities between the training and test splits. Thus, evaluation measures whether imputed training images support detection on ground-truth target-survey observations. These ground-truth images belong to the simulated benchmark; this protocol is separate from the future real-survey validation discussed in Section~\ref{sec:limitation}.

The approximately balanced test set is useful for controlled comparisons but does not represent the much lower lens prevalence expected in deployment. We therefore report prior-adjusted AUPRC for lens prevalence $\pi\in\{0.1,0.05,0.01,0.001\}$ in Table~\ref{tab:prior-adjusted-auprc}. Under a change in class prevalence with fixed class-conditional score distributions, AUROC is invariant, while precision at threshold $t$ becomes
\[
\operatorname{Precision}_{\pi}(t)=
\frac{\pi\,\operatorname{TPR}(t)}
{\pi\,\operatorname{TPR}(t)+(1-\pi)\,\operatorname{FPR}(t)}.
\]
This adjustment addresses prevalence shift without modeling additional changes in the lens or non-lens populations.

Absolute AUPRC decreases sharply as the assumed lens prevalence falls. Asyn-Diff remains the best-performing method and RePaint remains the lowest-performing method at every reported prior. The intermediate ordering changes. The limited number of test negatives constrains the reliability of the low-false-positive-rate region, so estimates become less reliable at very small prevalence. We therefore do not report priors below $\pi=0.001$.
\begin{table}[t]
\centering\small
\caption{Prior-adjusted AUPRC at target lens prevalence $\pi$ (higher is better).  Bold indicates the highest AUPRC in each column.}
\label{tab:prior-adjusted-auprc}
\begin{tabular}{lrrrr}
\toprule
Method & $\pi=0.1$ & $\pi=0.05$ & $\pi=0.01$ & $\pi=0.001$ \\
\midrule
Asyn-Diff & \textbf{0.6104} & \textbf{0.4747} & \textbf{0.2141} & \textbf{0.0411} \\
Palette & 0.5630 & 0.4180 & 0.1629 & 0.0339 \\
M2DN & 0.5388 & 0.4038 & 0.1750 & 0.0397 \\
RePaint & 0.5027 & 0.3546 & 0.1285 & 0.0288 \\
\bottomrule
\end{tabular}
\end{table}

\section{Angular Extent Error}
\label{app:angular-extent-error}

AEE is a secondary, lens-specific morphology diagnostic in our metric hierarchy. It measures recovery of lens-arc angular extent and complements the general reconstruction metrics and downstream lens classification. Table~\ref{tab:aee-summary} reports the task-averaged results; Table~\ref{tab:lens_flux_results_full} provides individual task results.
\begin{table}[t]
\centering
\small
\caption{Mean Angular Extent Error (AEE, radians; lower is better) over the three Roman-imputation tasks.  Task-level results are reported in Table~\ref{tab:lens_flux_results_full}. Best and second-best values are bold and underlined, respectively.}
\label{tab:aee-summary}
\begin{tabular}{lc}
\toprule
Method & AEE$\downarrow$ \\
\midrule
PSF-Linear & 1.426 \\
S\'ersic-FM & 1.460 \\
CodeBrain~\cite{wu2026codebrain}  & 1.424 \\
m2dn~\cite{m2dn} & 0.822 \\
RePaint~\cite{repaint} & 0.848 \\
Palette~\cite{saharia2022palette} & \second{0.811} \\
asyn-diff & \textbf{0.780} \\
\bottomrule
\end{tabular}
\end{table}

Angular Extent Error measures how well an imputed Roman image preserves the
azimuthal coverage of the strong-lensing arc.  The metric is computed from
binary arc masks rather than from all image pixels.  First, the Roman image is
collapsed to a single 2D signal map by clipping negative fluxes to zero and
averaging over the Roman bands.  To estimate the lens center, we apply a
Gaussian filter with $\sigma=1$ pixel to the deflector-only Roman image, take
the 99th percentile of the smoothed finite pixels as a high-flux threshold, and
compute the weighted centroid of pixels above that threshold using weights
equal to the threshold-subtracted flux.  If the image contains no positive
finite signal, the center defaults to the geometric image center; if the
thresholded region has zero total weight, the center defaults to the peak
smoothed pixel.

We then extract a ground-truth arc mask from the clean lensed-source Roman
image.  After the same 2D collapse and $\sigma=1$ smoothing, positive pixels
are thresholded at the larger of the 90th percentile of positive smoothed
values and 3\% of the positive peak value.  Connected components with fewer
than 6 pixels are discarded.  For a prediction, we first denormalize the
predicted Roman tensor back to flux space, collapse and smooth it in the same
way, and subtract a circularly symmetric lens-light profile: pixels are grouped
by integer radial distance from the estimated lens center, the median flux in
each radial bin is interpolated back to the image grid, and this radial profile
is subtracted.  Candidate predicted arc pixels are restricted to radii outside
$\max(1.5,0.35\,r_{\mathrm{gt}})$ pixels, where $r_{\mathrm{gt}}$ is the median
ground-truth arc radius in pixels when a valid ground-truth arc exists.  Within
this valid region, the residual threshold is the maximum of three quantities:
the residual median plus $3$ times the MAD-based robust standard deviation, the
97th percentile residual, and 3\% of the positive residual peak.  Components
with fewer than 6 pixels, or with median radius inside the allowed inner radius,
are removed; the default mask uses the union of all remaining components.

Given an arc mask $M$ and lens center $(c_x,c_y)$, we compute the polar angle
of every arc pixel,
$\theta_i=\operatorname{atan2}(y_i-c_y,x_i-c_x)$, wrap the angles to
$[0,2\pi)$, sort them, and find the largest gap between adjacent angles on the
circle, including the wrap-around gap.  The angular extent is
$\Delta\theta(M)=2\pi-\max_j(\theta_{j+1}-\theta_j)$, with a single-pixel arc
assigned extent 0.  The reported Angular Extent Error is the absolute
difference
$|\Delta\theta(M_{\mathrm{pred}})-\Delta\theta(M_{\mathrm{gt}})|$; the
relative version divides this value by
$|\Delta\theta(M_{\mathrm{gt}})|+\epsilon$ with $\epsilon=10^{-8}$.  Samples
without a valid ground-truth or predicted arc mask are marked invalid for this
metric.

\subsection{Reference Scale and Threshold Sensitivity}
\label{app:aee-sensitivity}
A PSF-Linear reference evaluated for this diagnostic obtains AEE of 1.457, compared with 0.754 for Asyn-Diff under the default extraction setting, a 48.2\% reduction. This reference value belongs to the present sensitivity study and is reported separately from the physical-baseline reconstruction results in Table~\ref{tab:lens_flux_results_full}.

The qualitative examples in Figure~\ref{fig:imputation-examples} show that recovered arcs can retain much of their extent while developing localized breaks. This helps explain why methods can have relatively close AEE values despite visible differences in arc continuity: angular extent measures a specific aspect of morphology rather than complete structural fidelity.

Table~\ref{tab:aee-sensitivity} evaluates nine extraction configurations in addition to the default, varying the ground-truth and prediction percentiles, prediction signal-to-noise threshold, and minimum component area. Absolute AEE values change with these choices, but Asyn-Diff retains the lowest AEE across all ten settings tested. This supports the stability of that comparison within the tested parameter range.
\begin{table}[t]
\centering\small
\caption{AEE sensitivity to arc-extraction settings (radians; lower is better) on LSST+Euclid$\rightarrow$Roman. Each alternative changes the indicated parameter from the default, keeping the other settings fixed. Bold indicates the lowest AEE in each row.}
\label{tab:aee-sensitivity}
\begin{adjustbox}{max width=\linewidth}
\begin{tabular}{lrrrr}
\toprule
Setting & Asyn-Diff & Palette & RePaint & M2DN \\
\midrule
Default & \textbf{0.754} & 0.790 & 0.786 & 0.781 \\
\texttt{gt\_percentile}: $90\rightarrow85$ & \textbf{0.763} & 0.804 & 0.801 & 0.789 \\
\texttt{gt\_percentile}: $90\rightarrow95$ & \textbf{0.663} & 0.690 & 0.692 & 0.688 \\
\texttt{pred\_percentile}: $97\rightarrow85$ & \textbf{0.648} & 0.658 & 0.708 & 0.651 \\
\texttt{pred\_percentile}: $97\rightarrow90$ & \textbf{0.647} & 0.657 & 0.708 & 0.651 \\
\texttt{pred\_percentile}: $97\rightarrow95$ & \textbf{0.673} & 0.686 & 0.733 & 0.695 \\
\texttt{pred\_snr}: $3\rightarrow2$ & \textbf{0.754} & 0.789 & 0.787 & 0.781 \\
\texttt{pred\_snr}: $3\rightarrow4$ & \textbf{0.755} & 0.792 & 0.789 & 0.782 \\
\texttt{min\_area}: $6\rightarrow3$ & \textbf{0.751} & 0.772 & 0.782 & 0.776 \\
\texttt{min\_area}: $6\rightarrow12$ & \textbf{0.770} & 0.814 & 0.801 & 0.795 \\
\bottomrule
\end{tabular}
\end{adjustbox}
\end{table}

\section{Few-Shot Mitigation of Rare-Event Shift}
\label{app:fewshot-mitigation}
The main-text rare-event imputation results evaluate zero-shot transfer from common galaxies to strong lenses. Here we investigate whether limited lens supervision can mitigate that shift for Asyn-Diff. We evaluate two strategies with 10, 20, 50, and 100 lens examples: common-galaxy pretraining followed by lens-only fine-tuning, and joint training with importance-weighted lens sampling. For joint training, we sample with replacement, assigning 25\% of the sampling probability mass to lenses and 75\% to common galaxies.

We run two trials for each strategy and each nonzero budget, using independently sampled few-shot lens subsets to assess sensitivity to which rare examples are selected. Tables~\ref{tab:fewshot-finetuning} and~\ref{tab:fewshot-joint-training} report lens-imputation CRPS. The repeated 0-shot value of 0.090 is the common-galaxy pretrained baseline, not an additional independently sampled few-shot trial.

\paragraph{Lens-only fine-tuning.}
Fine-tuning on the small lens subset increases CRPS above the 0-shot baseline in both trials at every tested budget. Thus, lens-only fine-tuning does not mitigate the rare-event shift under the tested settings.
\begin{table}[t]
\centering\small
\caption{Asyn-Diff lens-imputation CRPS after common-galaxy pretraining followed by lens-only fine-tuning (lower is better). Each trial uses an independently sampled few-shot lens subset. The 0-shot column repeats the shared pretrained baseline without lens supervision.}
\label{tab:fewshot-finetuning}
\begin{tabular}{lrrrrr}
\toprule
Trial & 0 shots & 10 shots & 20 shots & 50 shots & 100 shots \\
\midrule
Trial 1 & 0.090 & 0.136 & 0.124 & 0.125 & 0.124 \\
Trial 2 & 0.090 & 0.139 & 0.124 & 0.126 & 0.130 \\
\bottomrule
\end{tabular}
\end{table}

\paragraph{Importance-weighted joint training.}
Joint training can improve over the 0-shot baseline, but the gains are neither uniform across independently sampled subsets nor monotonic in the number of lens examples. At 100 shots, for example, Trial 1 reaches 0.085 while Trial 2 reaches 0.091, compared with the 0-shot value of 0.090. Limited rare-event supervision can therefore help, but its effectiveness depends strongly on sample composition. Two trials expose this sensitivity without establishing a precise estimate of its variability.
\begin{table}[t]
\centering\small
\caption{Asyn-Diff lens-imputation CRPS with importance-weighted joint training, implemented by weighted sampling with replacement: 25\% probability mass on lenses and 75\% on common galaxies (lower is better). Each trial uses an independently sampled lens subset. The 0-shot column repeats the shared common-galaxy baseline.}
\label{tab:fewshot-joint-training}
\begin{tabular}{lrrrrr}
\toprule
Trial & 0 shots & 10 shots & 20 shots & 50 shots & 100 shots \\
\midrule
Trial 1 & 0.090 & 0.091 & 0.091 & 0.085 & 0.085 \\
Trial 2 & 0.090 & 0.084 & 0.082 & 0.083 & 0.091 \\
\bottomrule
\end{tabular}
\end{table}

\paragraph{Interpretation and future directions.}
One plausible explanation is that the large common-galaxy set covers a broad range of galaxy appearances, whereas a small randomly selected lens subset may cover only a narrow, unrepresentative range of source and deflector properties. Promising directions include unsupervised or semi-supervised test-time adaptation using available LSST and Euclid observations of lens candidates.

\section{Detailed Tables}
\label{app:detailed-tables}

This section provides the complete results for all Roman-related imputation tasks.
The detailed results on common galaxies are shown in Table~\ref{tab:main_results_full},
and the detailed results on rare galaxy--galaxy strong lensing are shown in
Table~\ref{tab:lens_flux_results_full}. Downstream lens detection performance using different training data are shown in Table~\ref{tab:downstream-classification-auroc-auprc}

\providecommand{\second}[1]{\underline{#1}}

\begin{table}[t]
\caption{Comparison of cross-survey imputation methods under deterministic (RMSE, PSNR) and probabilistic (CRPS) metrics for Roman-target imputation tasks. Lower is better for RMSE and CRPS; higher is better for PSNR. Best results are in bold, and second-best results are underlined. CRPS is not reported for the two deterministic physical baselines. A slash denotes an unavailable or unreported result. }
\centering
\small
\label{tab:main_results_full}
\begin{tabular}{llccc}
\toprule
Task & Method & RMSE$\downarrow$ & PSNR$\uparrow$ & CRPS$\downarrow$ \\
\midrule

\multirow{11}{*}{Euclid $\rightarrow$ Roman}
& PSF-Linear & 0.256 & 36.33 & / \\
& S\'ersic-FM & 0.251 & 37.11 & / \\
& CodeBrain & 0.331 & 34.50 & 0.098 \\
& diffm4ri          & 0.361 & 33.68 & 0.101 \\
& mmgan             & 0.591 & 28.99 & 0.165 \\
& ummi              & 0.406 & 34.48 & 0.129 \\
& mmsyn             & 0.270 & 36.07 & 0.090 \\
& m2dn              & \second{0.219} & \second{38.10} & 0.077 \\
& RePaint        & 0.396 & 33.97 & 0.104 \\
& Palette         & 0.215 & 37.93 & \second{0.074} \\
& asyn-diff         & \textbf{0.183} & \textbf{39.36} & \textbf{0.070} \\

\midrule

\multirow{11}{*}{Euclid+LSST $\rightarrow$ Roman}
& PSF-Linear & 0.514 & 32.51 & / \\
& S\'ersic-FM & 0.267 & 37.47 & / \\
& CodeBrain & 0.314 & 34.88 & 0.096 \\
& diffm4ri          & 0.338 & 34.59 & 0.087 \\
& mmgan             & 0.455 & 31.29 & 0.152 \\
& ummi              & 0.402 & 34.48 & 0.127 \\
& mmsyn             & 0.265 & 36.17 & 0.091 \\
& m2dn              & 0.218 & 37.79 & 0.072 \\
& RePaint        & 0.232 & 37.53 & 0.073 \\
& Palette         & \second{0.160} & \second{40.17} & \second{0.069} \\
& asyn-diff         & \textbf{0.157} & \textbf{40.22} & \textbf{0.066} \\

\midrule

\multirow{11}{*}{LSST $\rightarrow$ Roman}
& PSF-Linear & 0.440 & 32.43 & / \\
& S\'ersic-FM & 0.295 & 35.94 & / \\
& CodeBrain & 0.343 & 34.22 & 0.098 \\
& diffm4ri          & 0.360 & 34.19 & 0.093 \\
& mmgan             & 0.519 & 30.19 & 0.160 \\
& ummi              & 0.540 & 31.74 & 0.145 \\
& mmsyn             & 0.312 & 34.94 & 0.095 \\
& m2dn              & 0.277 & 36.38 & 0.082 \\
& RePaint       & 0.337 & 35.02 & 0.093 \\
& Palette        & \second{0.212} & \textbf{38.11} & \second{0.074} \\
& asyn-diff         & \textbf{0.212} & \second{37.99} & \textbf{0.070} \\

\bottomrule
\end{tabular}
\end{table}
\providecommand{\second}[1]{\underline{#1}}

\begin{table}[t]

\caption{Performance comparison of cross-survey imputation methods on rare galaxy-galaxy strong lensing.
Results are reported for individual imputation tasks under distribution shift, where models trained on common galaxies are evaluated on rare events. Best results are in bold, and second-best results are underlined. CRPS is not reported for the two deterministic physical baselines. }
\label{tab:lens_flux_results_full}
\centering
\small
\begin{tabular}{llcccc}
\toprule
Task & Method & RMSE$\downarrow$ & PSNR$\uparrow$ & CRPS$\downarrow$ & AEE$\downarrow$\\
\midrule

\multirow{7}{*}{Euclid+LSST $\rightarrow$ Roman}
& PSF-Linear & 0.715 & 31.88 & / & 1.595 \\
& S\'ersic-FM & 0.565 & 34.58 & / & 1.495 \\
& CodeBrain & 0.413 & 34.56 & 0.119 & 1.400 \\
& m2dn      & 0.308 & 37.53 & 0.098 & \second{0.781} \\
& RePaint   & 0.362 & 36.29 & 0.101 & 0.786 \\
& Palette   & \second{0.235} & \second{39.05} & \second{0.085} & 0.790 \\
& asyn-diff & \textbf{0.234} & \textbf{39.06} & \textbf{0.079} & \textbf{0.754} \\

\midrule

\multirow{7}{*}{LSST $\rightarrow$ Roman}
& PSF-Linear & 0.599 & 31.84 & / & 1.613 \\
& S\'ersic-FM & 0.551 & 33.05 & / & 1.520 \\
& CodeBrain & 0.462 & 33.76 & 0.126 & 1.484 \\
& m2dn      & 0.385 & 35.83 & 0.111 & 0.845 \\
& RePaint   & 0.537 & 33.47 & 0.148 & 0.954 \\
& Palette   & \second{0.315} & \second{36.74} & \second{0.092} & \second{0.821} \\
& asyn-diff & \textbf{0.309} & \textbf{36.82} & \textbf{0.088} & \textbf{0.813} \\

\midrule

\multirow{7}{*}{Euclid $\rightarrow$ Roman}
& PSF-Linear & 0.400 & 34.96 & / & 1.071 \\
& S\'ersic-FM & 0.499 & 33.65 & / & 1.365 \\
& CodeBrain & 0.444 & 34.06 & 0.126 & 1.388 \\
& m2dn      & \second{0.374} & \second{35.72} & \second{0.108} & 0.839 \\
& RePaint   & 0.668 & 32.36 & 0.182 & 0.803 \\
& Palette   & 0.381 & 35.70 & 0.107 & \second{0.821} \\
& asyn-diff & \textbf{0.347} & \textbf{36.70} & \textbf{0.104} & \textbf{0.773} \\

\bottomrule
\end{tabular}

\end{table}
\begin{table}[t]
\centering
\caption{Downstream Roman lens detection performance.}
\label{tab:downstream-classification-auroc-auprc}
\begin{tabular}{lllcc}
\toprule
Train source & Task & Method & AUROC & AUPRC \\
\midrule
GT Roman & -- & -- & 0.9626 & 0.9606 \\
\midrule
Imputed Roman & LSST+Euclid $\rightarrow$ Roman & asyn-diffusion & 0.9056 & 0.8978 \\
 &  & RePaint & 0.8782 & 0.8623 \\
 &  & Palette & 0.8934 & 0.8832 \\
 &  & m2dn & 0.8794 & 0.8688 \\
 & & PSF-Linear & 0.7892 & 0.7662 \\
 & & S\'ersic-FM & 0.6882 & 0.6181 \\
 & & CodeBrain  & 0.8353 & 0.8176 \\
 & LSST $\rightarrow$ Roman & asyn-diffusion & 0.8663 & 0.8495 \\
 &  & RePaint & 0.7907 & 0.7451 \\
 &  & Palette & 0.8575 & 0.8377 \\
 &  & m2dn & 0.8487 & 0.8439 \\
 & & PSF-Linear & 0.6778 & 0.6414 \\
 & & S\'ersic-FM & 0.7214 & 0.6274 \\
 & & CodeBrain  & 0.8083 & 0.7940 \\
 & Euclid $\rightarrow$ Roman & asyn-diffusion & 0.8881 & 0.8887 \\
 &  & RePaint & 0.7619 & 0.7421 \\
 &  & Palette & 0.8928 & 0.8857 \\
 &  & m2dn & 0.8700 & 0.8624 \\
 & & PSF-Linear & 0.8938 & 0.8923 \\
 & & S\'ersic-FM & 0.7953 & 0.6942 \\
 & & CodeBrain & 0.8124 & 0.7920 \\
\bottomrule
\end{tabular}
\end{table}

\section{Photometric and Morphological Diagnostics}
\label{app:scientific-diagnostics}
Table~\ref{tab:scientific-diagnostics} reports RMSEs for total flux, aperture flux, ellipticity, and second-moment size $T$. Asyn-Diff has the lowest size error and ties M2DN for the lowest ellipticity error. Palette has the lowest aperture-flux error, while S\'ersic-FM has the lowest total-flux error. These results show that the ranking depends on the scientific statistic: accurate total-flux recovery does not by itself imply accurate morphology or strong downstream detection performance.

These measurements serve as complementary diagnostics. COSMIO targets spatially resolved information, including irregular structures, that cannot be summarized by a few scalar quantities. If a particular statistic is the sole scientific target, a dedicated estimator with well-understood statistical properties is preferable. The superior total-flux result of S\'ersic-FM illustrates this distinction; it does not contradict the motivation for evaluating image-level generative imputation through morphology and rare-event detection.
\begin{table}[t]
\centering
\small
\caption{RMSE of photometric and low-order morphological statistics. Lower is better within each row; bold indicates the lowest error, including ties.}
\label{tab:scientific-diagnostics}
\begin{tabular}{lrrrr}
\toprule
Statistic & Asyn-Diff & Palette & M2DN & S\'ersic-FM \\
\midrule
Total flux & 139.30 & 345.70 & 479.50 & \textbf{83.57} \\
Aperture flux & 41.39 & \textbf{32.95} & 77.25 & 69.97 \\
Ellipticity & \textbf{0.054} & 0.060 & \textbf{0.054} & 0.210 \\
Second-moment size $T$ & \textbf{0.84} & 0.90 & 1.30 & 3.50 \\
\bottomrule
\end{tabular}
\end{table}

\section{Additional Dataset Visualizations}
\label{app:dataset_visualizations}

\begin{figure}[t]
    \centering
    \includegraphics[width=\textwidth]{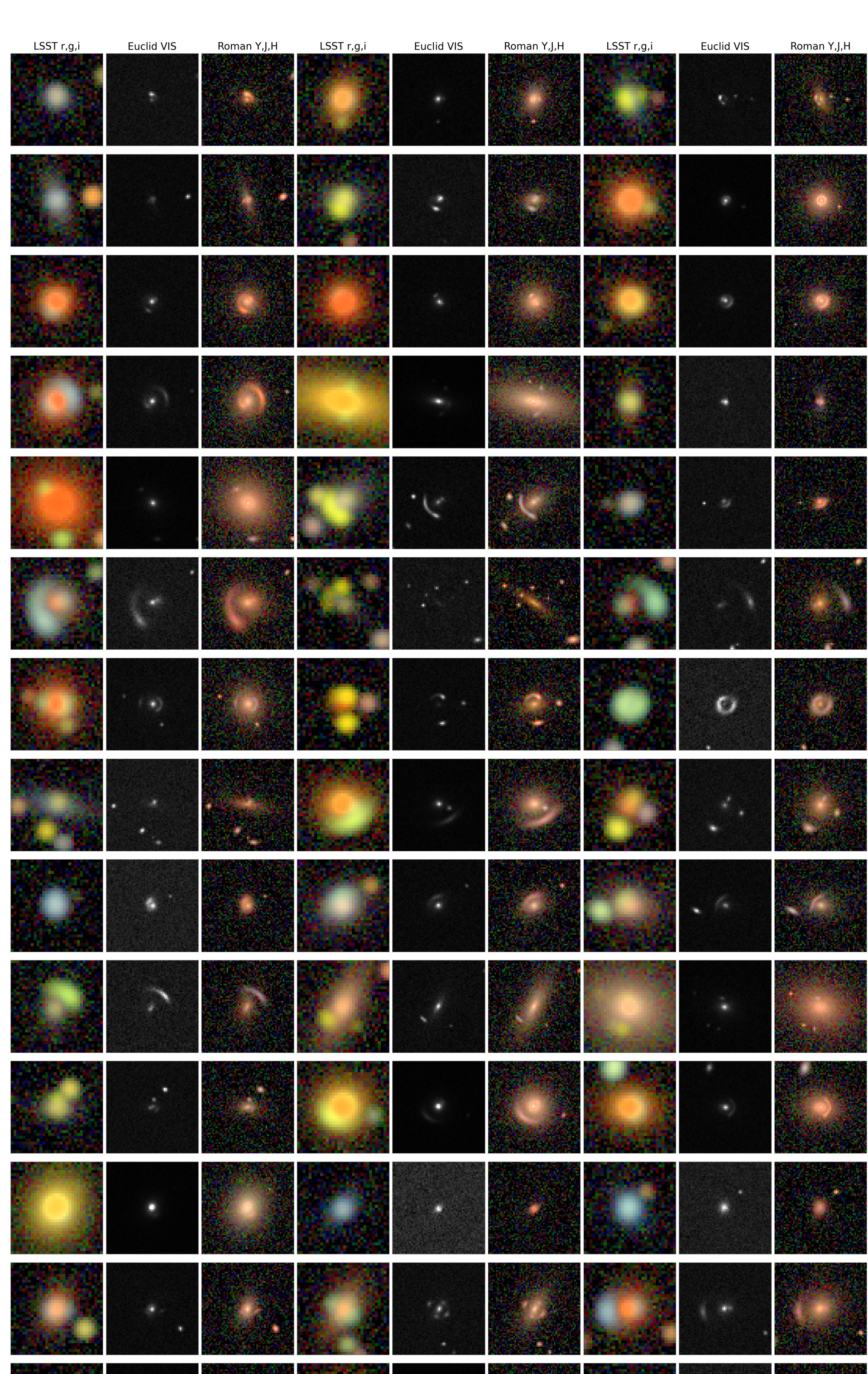}
    \caption{Examples of galaxy--galaxy strong-lensing systems in the COSMIO dataset.}
    \label{fig:lens_data_examples}
\end{figure}

\begin{figure*}[t]
    \centering
    \includegraphics[width=\textwidth]{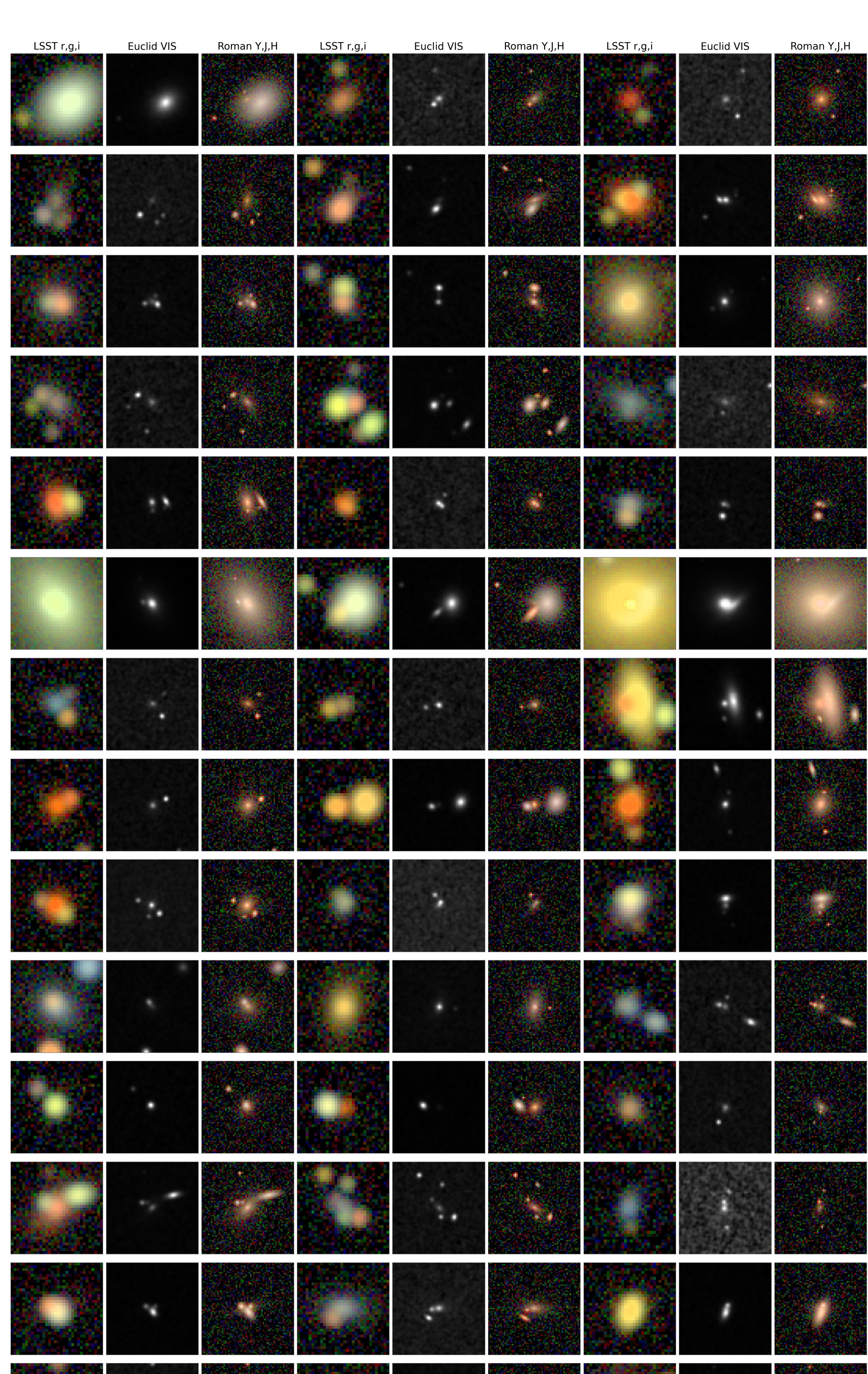}
    \caption{Examples of regular galaxies in the COSMIO dataset.}
    \label{fig:nonlens_data_examples}
\end{figure*}

Additional visualizations of the COSMIO dataset are shown in
Figure~\ref{fig:lens_data_examples} for strong lensing and Figure~\ref{fig:nonlens_data_examples} for regular galaxies.

\section{Additional Imputation Visualizations}
\label{app:imputation_visualizations}
\begin{figure}[t]
    \centering
    \includegraphics[width=\textwidth]{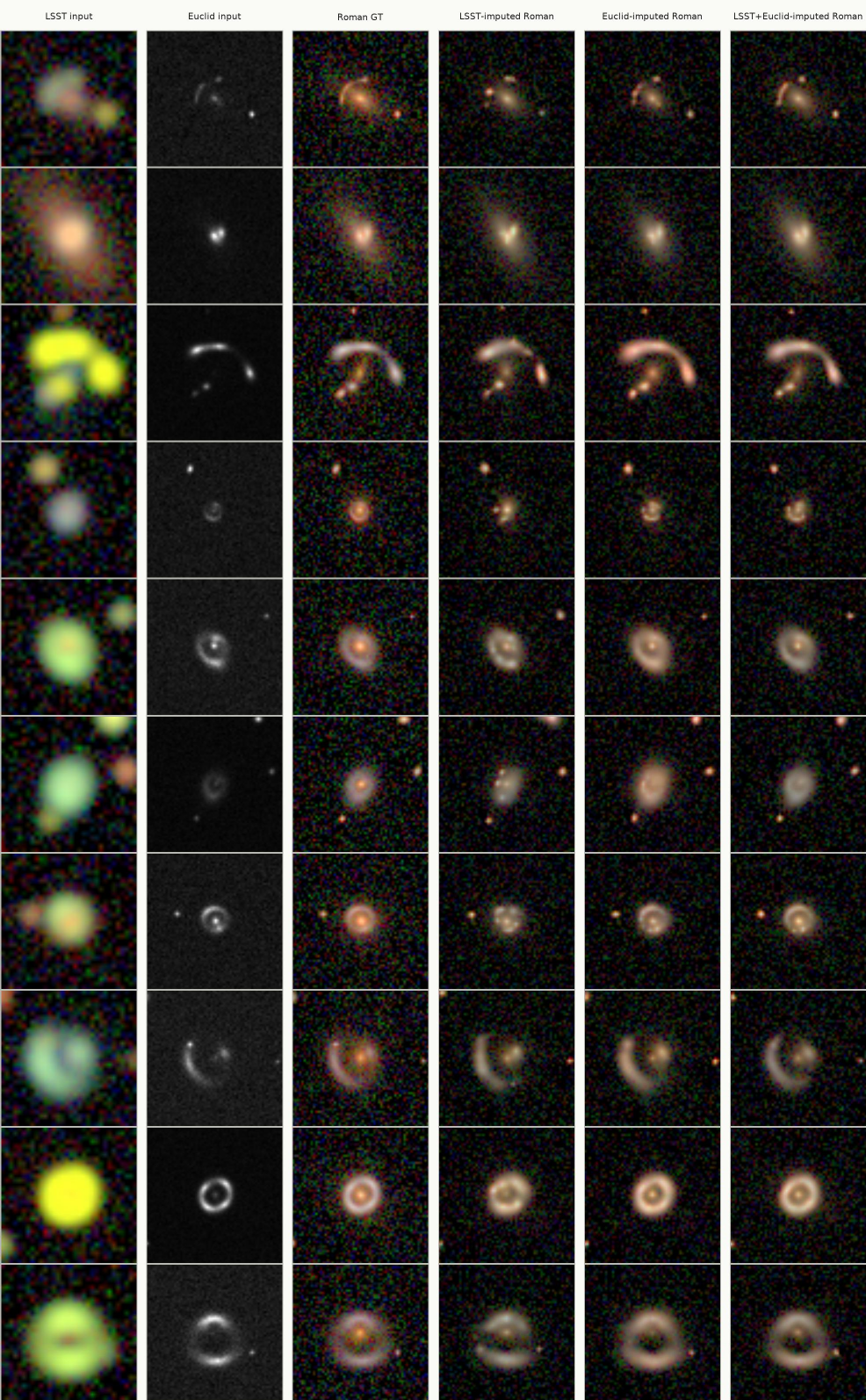}
    \caption{Additional qualitative examples of strong-lensing imputation using asynchronous diffusion.}
    \label{fig:lens_asyn_diff}
\end{figure}

\begin{figure}[t]
    \centering
    \includegraphics[width=\textwidth]{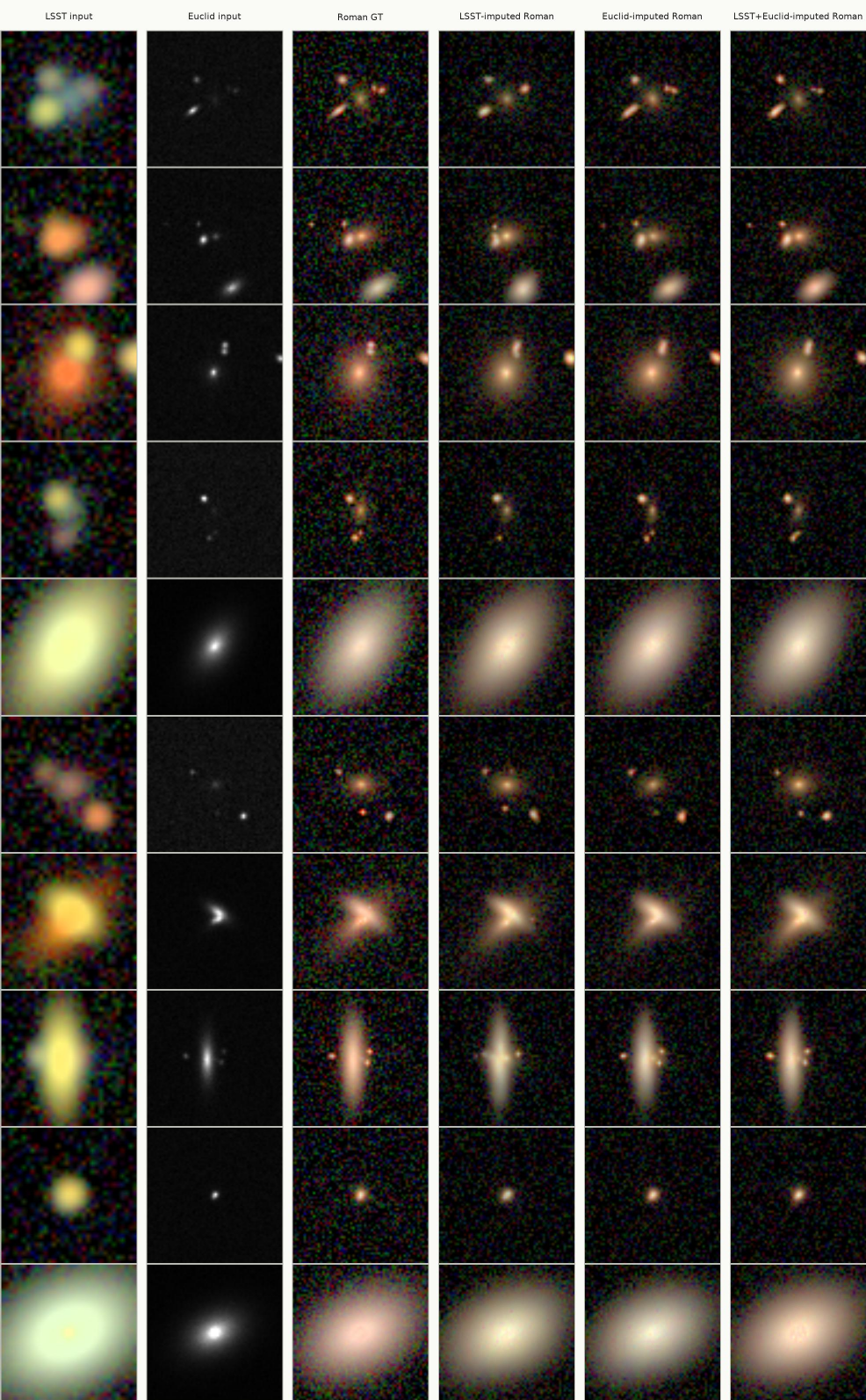}
    \caption{Additional qualitative examples of regular-galaxy imputation using asynchronous diffusion.}
    \label{fig:nonlens_asyn_diff}
\end{figure}

Additional visualization of imputed Roman Observations with asynchronous diffusion are shown in Figure~\ref{fig:lens_asyn_diff} for strong lensing, and Figure.~\ref{fig:nonlens_asyn_diff} for regular galaxies.



\newpage
\clearpage
\section*{NeurIPS Paper Checklist}

\begin{enumerate}

\item {\bf Claims}
    \item[] Question: Do the main claims made in the abstract and introduction accurately reflect the paper's contributions and scope?
    \item[] Answer: \answerYes{} 
    \item[] Justification: The claim made by Abstract and Introduction are reflected in the experimental results in Section.~\ref{sec:results}.

\item {\bf Limitations}
    \item[] Question: Does the paper discuss the limitations of the work performed by the authors?
    \item[] Answer: \answerYes{} 
    \item[] Justification: We use Section~\ref{sec:limitation} to discuss the limitation of this paper and benchmark.


\item {\bf Theory assumptions and proofs}
    \item[] Question: For each theoretical result, does the paper provide the full set of assumptions and a complete (and correct) proof?
    \item[] Answer: \answerNA{} 
    \item[] Justification: The paper does not include theoretical results, formal theorems, lemmas, or proofs. It is an empirical benchmark paper focused on dataset construction, experimental evaluation, and downstream analysis.

    \item {\bf Experimental result reproducibility}
    \item[] Question: Does the paper fully disclose all the information needed to reproduce the main experimental results of the paper to the extent that it affects the main claims and/or conclusions of the paper (regardless of whether the code and data are provided or not)?
    \item[] Answer: \answerYes{} 
    \item[] Justification: All experimental setups are in Section.~\ref{sec:experimental-setup}. All architecture details are in Appendix~\ref{app:training-details}. Release code and data also contains everything include config files for reproduction.

\item {\bf Open access to data and code}
    \item[] Question: Does the paper provide open access to the data and code, with sufficient instructions to faithfully reproduce the main experimental results, as described in supplemental material?
    \item[] Answer: \answerYes{} 
    \item[] Justification:  We provide open access to the code and data through the repository linked in the paper.

\item {\bf Experimental setting/details}
    \item[] Question: Does the paper specify all the training and test details (e.g., data splits, hyperparameters, how they were chosen, type of optimizer) necessary to understand the results?
    \item[] Answer: \answerYes{} 
    \item[] Justification: All experimental setups are in Section.~\ref{sec:experimental-setup}.

\item {\bf Experiment statistical significance}
    \item[] Question: Does the paper report error bars suitably and correctly defined or other appropriate information about the statistical significance of the experiments?
    \item[] Answer: \answerNo{} 
    \item[] Justification: The benchmark evaluates a broad set of methods across multiple imputation settings, target modalities, and downstream tasks, repeating every method under multiple random seeds is computationally prohibitive. Instead, the main empirical claims are supported by consistent trends across different model families, imputation patterns, and evaluation metrics, rather than by multiple repetition of a single experiment.

\item {\bf Experiments compute resources}
    \item[] Question: For each experiment, does the paper provide sufficient information on the computer resources (type of compute workers, memory, time of execution) needed to reproduce the experiments?
    \item[] Answer: \answerYes{} 
    \item[] Justification: We report in Section.~\ref{sec:experimental-setup}.
    
\item {\bf Code of ethics}
    \item[] Question: Does the research conducted in the paper conform, in every respect, with the NeurIPS Code of Ethics \url{https://neurips.cc/public/EthicsGuidelines}?
    \item[] Answer: \answerYes{} 
    \item[] Justification: We have read and understood the code of ethics and have made every effort
to adhere to it.

\item {\bf Broader impacts}
    \item[] Question: Does the paper discuss both potential positive societal impacts and negative societal impacts of the work performed?
    \item[] Answer: \answerNA{} 
    \item[] Justification: This work concerns a benchmark for observational cosmology data. It is intended for methodological development and scientific analysis in astronomy, and does not involve human subjects, social groups, or deployment-facing decision systems. Therefore, broader societal impacts are not directly applicable.
    
\item {\bf Safeguards}
    \item[] Question: Does the paper describe safeguards that have been put in place for responsible release of data or models that have a high risk for misuse (e.g., pre-trained language models, image generators, or scraped datasets)?
    \item[] Answer: \answerNA{} 
    \item[] Justification: This work releases observational cosmology data and benchmark code for scientific imputation experiments. The released data do not contain human-subject information, private data, scraped personal content, or deployment-facing generative models with high misuse potential. Therefore, the work does not pose the type of release risk targeted by this safeguard question

\item {\bf Licenses for existing assets}
    \item[] Question: Are the creators or original owners of assets (e.g., code, data, models), used in the paper, properly credited and are the license and terms of use explicitly mentioned and properly respected?
    \item[] Answer: \answerYes{} 
    \item[] Justification: The software packages used for forward rendering the survey scenes are credited in Section~\ref{sec:cosmio}. We cite the original sources for the simulation and rendering tools used in constructing the benchmark, and we use these assets consistently with their stated licenses and terms of use.

\item {\bf New assets}
    \item[] Question: Are new assets introduced in the paper well documented and is the documentation provided alongside the assets?
    \item[] Answer: \answerYes{} 
    \item[] Justification: The paper introduces and releases a new benchmark dataset and associated code. The dataset construction process, survey modalities, simulation pipeline, evaluation protocols, and limitations are provided. The released  repository provides the dataset files, code, configuration files, and instructions needed to use the assets and reproduce the main experiments.

\item {\bf Crowdsourcing and research with human subjects}
    \item[] Question: For crowdsourcing experiments and research with human subjects, does the paper include the full text of instructions given to participants and screenshots, if applicable, as well as details about compensation (if any)? 
    \item[] Answer: \answerNA{} 
    \item[] Justification: The paper does not involve crowdsourcing nor research with human subjects.

\item {\bf Institutional review board (IRB) approvals or equivalent for research with human subjects}
    \item[] Question: Does the paper describe potential risks incurred by study participants, whether such risks were disclosed to the subjects, and whether Institutional Review Board (IRB) approvals (or an equivalent approval/review based on the requirements of your country or institution) were obtained?
    \item[] Answer: \answerNA{} 
    \item[] Justification: The paper does not involve crowdsourcing nor research with human subjects.

\item {\bf Declaration of LLM usage}
    \item[] Question: Does the paper describe the usage of LLMs if it is an important, original, or non-standard component of the core methods in this research? Note that if the LLM is used only for writing, editing, or formatting purposes and does \emph{not} impact the core methodology, scientific rigor, or originality of the research, declaration is not required.
    \item[] Answer: \answerNA{} 
    \item[] Justification: LLMs are not used as part of the benchmark construction and simulation pipeline.

\end{enumerate}

\end{document}